\documentclass{aa}  

\usepackage{color}
\usepackage{natbib}
\usepackage{natbib,twoopt}
\usepackage[breaklinks=true]{hyperref} 
\usepackage{booktabs}
\usepackage{multirow}
\bibpunct{(}{)}{;}{a}{}{,} 
\usepackage{graphicx}
\newcommand{\ra}[1]{\renewcommand{\arraystretch}{#1}}
\usepackage[varg]{txfonts}
\begin{document}

\title{Consistent metallicity scale for cool dwarfs and giants \thanks{Based 
on data obtained from the ESO Science Archive Facility. The observations were made with ESO Telescopes at the La Silla and Paranal Observatories under programmes ID 070.D-0356, 088.C-0513 and 070.D-0421.}}

 \subtitle{A benchmark test using the Hyades}

   \author{Dutra-Ferreira, L.
          \inst{1,2}\fnmsep\thanks{\email{leticia@dfte.ufrn.br}}\,\,\,\,
          \and
          Pasquini, L.\inst{3}
          \and
          Smiljanic, R.\inst{4}
          \and
          Porto de Mello, G. F.,\inst{2} 
          \and Steffen, M. \inst{5} 
          }

  \institute{Departamento de F\'isica Te\'orica e Experimental, Universidade Federal do Rio Grande do Norte, Campus Universit\'ario Lagoa Nova, 59072-970 Natal, RN, Brazil\\           
         \and
             Universidade do Federal do Rio de Janeiro, Observat\'orio do Valongo, Ladeira do Pedro Ant\^onio 43, Rio de Janeiro, RJ, 20080-090, Brazil
          \and
             European Southern Observatory, 85748 Garching bei M\"unchen, Germany
         \and Department for Astrophysics, Nicolaus Copernicus Astronomical Center, ul. Rabiańska 8, 87-100 Toruń, Poland
         \and Leibniz-Institut f\"ur Astrophysik Potsdam, An der Sternwarte 16, D-14482 Potsdam, Germany\\
             }

   \date{Received; accepted }

 
  \abstract
   {In several instances chemical abundances of main-sequence and giant stars are used simultaneously under the assumption that they share the same abundance scale. This assumption, if wrong, might have important implications in different astrophysical contexts.}
   {It is therefore crucial to understand whether the metallicity or abundance differences among dwarfs and giants are real or are produced by systematic errors in the analysis. We aim to ascertain a methodology capable of producing a consistent metallicity scale for giants and dwarfs.}
  {To achieve that, we analyzed giants and dwarfs in the Hyades open cluster, under the assumption that they share the same chemical composition. All the stars in this cluster have archival high-resolution spectroscopic data obtained with HARPS and UVES. In addition, the giants have interferometric measurements of the angular diameters. We analyzed the sample with two methods. The first method constrains the atmospheric parameters independently from spectroscopic method. For that we present a novel calibration of microturbulence based on 3D model atmospheres. The second method is the classical spectroscopic analysis based on \ion{Fe}{}lines. We  also tested two different line lists in an attempt to minimize possible non-LTE effects and to optimize the treatment of the giants.}
  {We show that it is possible to obtain a consistent metallicity scale between dwarfs and giants. The preferred method should constrain the three parameters  $T_{\rm eff}$, $\log~g$, and $\xi$ independent of spectroscopy. A careful selection of \ion{Fe}{}~lines is also important. In particular, the lines should not be chosen based on the Sun or other dwarfs, but specifically to be free of blends in the spectra of giants. When attention is paid to the line list, the classical spectroscopic method can also produce consistent results. In our test, the metallicities derived with the well-constrained set of stellar parameters are consistent independent of the line list used. Therefore, for this cluster we favor the metallicity of +0.18$\pm$0.03~dex obtained with this method. The classical spectroscopic analysis, using the line list optimized for the giants, provides a metallicity of +0.14$\pm$0.03~dex, in agreement with previous works.}
   {}

   \maketitle
%

\section{Introduction}

The determination of chemical abundances in stars by spectral synthesis or 
curve of growth is a rather well-established technique, largely available to most 
astronomers. A more detailed view of stellar abundances has become possible 
since the advent of 8-10m class telescopes coupled with high-efficiency 
spectrographs, which produced high quality spectra for many stars. As a consequence, 
uncertainties in abundance analyses are mainly dominated by systematic errors rather 
than by observational limitations regarding spectral resolution and/or signal-to-noise 
ratio (S/N). In this context, we should recall that the determination of stellar 
abundances requires a number of assumptions, some of which might not be valid for 
stars in different evolutionary stages, such as dwarfs and giants. 

Although cool dwarfs and giants are often analyzed following a single 
methodology, the differences in their photospheric properties might introduce 
distinct systematic effects on the final atmospheric parameters derived during 
the analysis. An example is the large experiment with multiple analysis 
pipelines conducted within the Gaia-ESO Spectroscopic Survey. In this survey, 
it was clearly demonstrated that the multiple analysis methodologies perform 
differently in distinct regions of the parameter space 
\citep[see][]{Smiljanic14}. It is thus important to understand if differences 
on metallicities and abundances between giants and dwarfs are real or produced 
by systematic errors in the analysis. The interpretation of some open 
questions in astronomy depends on these kinds of comparisons.

For instance, metallicity plays a role in one of the most accepted scenarios 
in planet formation theories, which is the core accretion scenario. 
\citep[see, e.g.,][]{Pollack96}. Concerning the 
gas-giant planet formation, many works have shown that main-sequence FGK-type 
stars hosting giant planets ($>$~1.0 $\rm M_{J}$) are usually 
metal-rich when compared with their counterparts without giant planets 
\citep{Gonzalez97,Santos04,FisVal05a}. However,  a giant 
planet versus metallicity correlation is not well established for evolved stars. 
\cite{Pasquini07} showed that giant stars with planets do not tend to have 
high metallicities. A similar result was found by other works 
\citep{Schuler05,Takeda08,Ghezzi10,Mortier13}, but not by 
\cite{HekkerMelen07,Reffert15} who also argued in favor of a planet metallicity 
correlation  in giants. 

The situation for giants is more complex because of many factors as, for 
example, the limited sample size of giants hosting planets, inhomogeneities 
in the planetary properties, and the higher masses of  giant stars in comparison 
to dwarfs. Indeed, there is a stellar mass vs. planet vs. metallicity correlation 
when we consider the giant planet occurrence frequency in evolved stars. Some 
authors have reported that there is a trend toward higher metallicities for stars 
with masses above 1.5$M_{\odot}$ \citep{Johnson10,Maldonado13}. Understanding the 
metallicity distribution of stars with planets would provide an important constraint 
for planet formation theories.

The comparison between the metallicity scale of giants and dwarfs is also 
important in studies of Galactic chemical evolution. For instance, the 
metallicity distribution (MD) of the Galactic bulge provides clues about how 
similar bulge stars are to thin and thick disk stars. As a consequence, it 
aids the determination of the bulge age. Initial studies of the bulge were 
focused on giant stars \citep[e.g.,][]{McWilliam&Rich94,Zoccali06} because 
these are intrinsically brighter objects. Later, the microlensing technique 
was used to observe dwarfs and subgiants in the Galactic bulge. This revealed 
discrepancies between the MD of bulge giants and dwarfs 
\citep{Cohen08,Bensby10,Bensby11}. More recent studies, however, show a better 
agreement between the two MDs \citep{Bensby13,Anders14}, although the authors 
recognize some evidence of a bias toward the high-metallicity tail of the 
giants' distribution \citep{Taylor&Croxall05,Hill11}. Indeed, it is difficult 
to analyze giants in the high-metallicity regime since their spectra are severely affected by blending and  molecules 
features because of their
cool 
atmospheres. A better picture of the MD of giants requires a full knowledge of the 
source of this bias in the metal-rich regime. 
 
Methodological limitations may affect  the analysis of giants and 
dwarfs differently, e.g., how realistic are the adopted atmospheric models, continuum 
normalization, and  atomic and molecular constants used. The use of the 
same line list can be a challenge since the intensity of the spectral lines 
is different in these objects. Moreover, departures from the local 
thermodynamical equilibrium (LTE) are particularly important for low-gravity 
and low-metallicity stars \citep{Asplund05}. Spectroscopic gravities derived 
by the ionization equilibrium may be unreliable because of \ion{Fe}{i} 
overionization. High-temperature dwarfs ($T_{\rm{eff}}$~$>$~6000~K) may also 
suffer from non-LTE effects \citep{Mashon10}. Additionally, the 
\emph{gf} values need to be very accurate for giants since a differential 
analysis with respect to the Sun does not cancel out uncertainties in these 
constants.

Open clusters are the optimal sites to evaluate the limitations of abundance 
analyses when applied to giants and dwarfs. They are made of stars with 
basically the same distance, and in general, it is reasonable to assume that 
all stars share the same overall chemical composition, except for the elements 
affected by mixing in giants 
\citep[see, e.g.,][]{Takeda08,Smiljanic09,Verne13}. Stars in an open cluster 
also share the same age, outlining a common isochrone curve in the HR diagram, 
making it easier to constrain stellar parameters in these environments. 

Few studies  so far have attempted the simultaneous analysis of giants and 
dwarfs in open clusters, aiming to explore possible discrepancies in the 
abundance patterns between these two classes of objects. \cite{Pasquini04} 
investigated dwarfs and giants of the intermediate-age cluster IC~4651, and 
found, in general, excellent agreement between stars of different evolutionary 
status. \cite{Pace2010} investigated the metallicity of five dwarfs and three 
giants in two open clusters and found differences in the metallicity of up to 
0.10~dex for one of the clusters. They also pointed out enhancements of 
sodium, aluminium, and silicon for the giants. A similar study was performed 
by \citet{San2009,San2012}, who investigated the abundance pattern of several 
open clusters performing a simultaneous and homogeneous spectroscopic 
analysis. These authors noticed that the discrepancy on the metallicity scale 
of giants and dwarfs belonging to the same cluster may depend on the line list 
used. The explanation about the source of these differences is still a matter 
of investigation.

In this work, we chose the Hyades open cluster to perform a simultaneous and 
homogeneous study of giant and dwarf stars. Our aim is to define an analysis 
method that can deliver metallicities in a consistent scale for both types of 
stars. Once this method is tested and established, it can be applied to 
different astrophysical problems, such  as the comparison of MDs of 
planet host dwarfs and giants. We choose the Hyades, as atmospheric parameters 
can be constrained by other methods than the classical spectroscopic analysis 
(see below), and because this cluster is young enough that it should be free of 
atomic diffusion effects. The paper is organized as follows. Section \ref{sec:hyades} 
reviews the main properties of the cluster, while Section \ref{sec:data} 
presents a description of the data. Section \ref{sec:analysis} is dedicated to 
the analysis and Section \ref{sec:discussion} to the discussion of the results. 
In Section \ref{sec:conclusion} we draw our final conclusions.


\section{Hyades: The benchmark test} \label{sec:hyades}

The Hyades is a relatively young cluster, with an estimated age of 
$\sim$~625~$\pm$~50~$\rm{Myr}$ \citep{P98}, and is the closest open cluster 
to the Sun \citep[$\sim$~46.5~pc,][]{vanLeeuwen09}. Spectroscopic studies of 
FGK-type dwarfs in the Hyades find a metallicity of about +0.13~dex 
\citep{Cayrel85,Boesgaard90,Paulson03,Schuler06a}. Regarding the giants, the 
metallicity values range from +0.10 up to +0.20~dex, where this scatter is 
usually attributed to the star HIP~20455, a spectroscopic binary 
\citep{Schuler06a,CarPan11}.

The slightly over-solar metallicity provides a safe regime to test the 
classical spectroscopic analysis. At this regime, departures from LTE are not 
expected to be significant for \ion{Fe}{i}. Also, close to the solar 
metallicity, the mean temperature stratification is close to the radiative 
equilibrium expectation, and therefore, the difference between 1D and 3D model 
atmospheres is expected to be relatively small \citep{Asplund05}. 

A relevant advantage of the stars in the Hyades is that their atmospheric 
parameters can be very well constrained. From the standpoint of the giants, 
interferometric measurements of the angular diameters are available, which 
enables the \emph{\textup{direct}} determination of absolute effective temperatures 
\citep{Boyajian09}. Furthermore, precise Hipparcos parallaxes 
\citep{vanLeeuwen07} are available. Thus, a reliable determination of the 
surface gravities is possible. Finally, a large amount of spectroscopic data 
is available for the numerous dwarfs and all  four giants of this cluster. All 
these make the Hyades the optimal benchmark to test the limitations of the 
classical spectroscopic analysis method. 

The Hyades is also a target of planet searches using the radial velocity 
technique \citep{Cochran02,Paulson04}. So far, one of the four giants in the 
Hyades was reported to host a giant planet. The clump giant HIP~20889 has a 
long period planet ($\sim$ 594 days) with $\sim$~7.6~$M_{J}$ \citep{Sato07}. 
More recently, \cite{Quinn14} reported the discovery of the first hot Jupiter 
orbiting a K dwarf HD~285507 in this cluster. 

Our sample was selected as follows. The dwarfs were selected primarily from 
the list of \cite{vanBueren52} and then cross-checked with the reliable sample 
of cluster members defined by \cite{P98}. Three additional dwarfs were selected 
exclusively from \cite{P98} to complement the cooler end of our 
sample (these are the stars without the vB number in Table~\ref{tab1}). Among 
them, one cool dwarf (HIP~13976) appears slightly away from the cluster's main 
sequence. However, this object was classified as a reliable member of the 
cluster by \cite{P98} and has radial velocity and distance fully compatible 
with the cluster distribution. Moreover, this star is present in many analyses 
of the cluster \citep[e.g.,][]{Paulson03,Yong04,Schuler06a}. 

\begin{figure}
\centering
\includegraphics[width=0.45\textwidth]{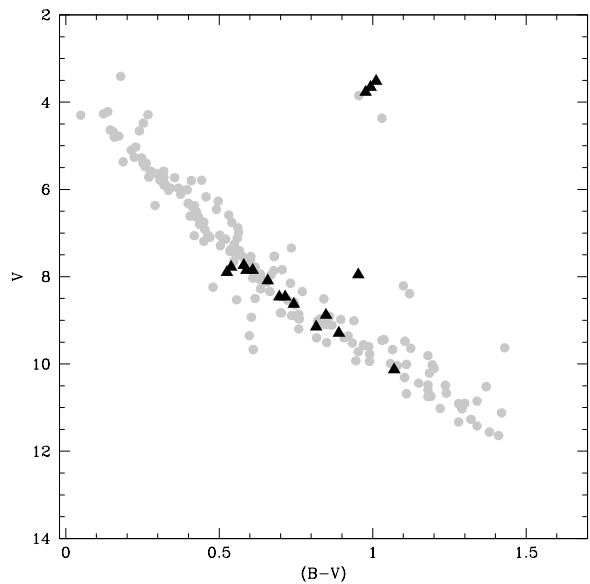}
 \caption{Color magnitude diagram (CMD) of the Hyades. The large black triangles correspond to the sample selected for our analysis. The small gray dots are the other cluster members. The magnitude $V$ and the $(B-V)$ color are from \cite{P98} and all stars shown fulfill the membership criterion of this work.}\label{fig1}
\end{figure}

Several studies investigated binaries in the Hyades 
\citep{Stefanik&Lathan85,Stefanik&Lathan92} and our stars were chosen to avoid 
binary systems. Among the giants, we excluded HIP~20885, as it is a 
spectroscopic binary (SB1) with a blue companion. \cite{Torres97} estimated 
that the secondary star contributes about 3\% of the flux of the primary. An 
accurate abundance analysis should take this contribution into account. 
Finally, whenever possible, we chose stars that have been studied in previous 
works for comparison purposes \citep[in particular,][]{Paulson03,Schuler06a}. 
Table~\ref{tab1} presents the basic data of the 14 dwarfs and the three giants 
selected for our sample. Their position in the CMD is shown in 
Fig.~\ref{fig1}. In this figure, the magnitude $V$ and the $(B-V)$ color are 
from \cite{P98}. We highlight that our sample encompasses stars in a large 
range of effective temperatures (4700~K~$\le$~$T_\mathrm{eff}$~$\le$~6200~K). 
This facilitates the investigation of possible systematic effects from the analysis 
as a function of this parameter. 

\begin{table*}
\ra{1.2}
\small
 \caption{Selected objects in the Hyades cluster. The letters in the columns refer to: (a) van Bueren number; (b) V magnitude from \cite{Johnson55}, except for the stars with number marks; (c) Hipparcos parallax and its respective standard error (mas); (d) Radial velocity from \cite{P98} and its respective error; (e) adopted mass in solar units (see Section \ref{sec:m1} for details).} 
 \label{tab1}
\centering
\begin{tabular}{cccccccccc}
\toprule
&   \multicolumn{8}{c}{Giants} \\
\midrule
HIP  &  vB  & Spec. Type &     V       & RA(J2000)  & Dec(J2000)  & ($\pi$~$\pm$~$\sigma_\pi$) & ($V_{r}$~$\pm$~$\sigma_{V_{r}}$) & $M_\sun$  & S/N \\
     &      &          &             &             &            &  (mas)         & ($\mathrm{km~s^{-1}}$) & &  @~609~nm  \\     & (a)  &           &    (b)      &             &            &   (c)          &   (d)                  & (e)&          \\ 
\hline
20205 &  28  & K0III    & 3.66         & 04:19:47.6  & +15:37:39.5 & 21.17$\pm$1.17        & +39.28$\pm$0.11        & 2.48 & 400 \\
20455 &  41  & K0IV     & 3.77         & 04:22:56.1  & +17:32:33.0 & 21.29$\pm$0.93        & +39.65$\pm$0.08        & 2.48 & 380 \\
20889 &  70  & K0III    & 3.52         & 04:28:36.9  & +19:10:49.5 & 21.04$\pm$0.82        & +39.37$\pm$0.06        & 2.48 & 440 \\
\midrule
 &   \multicolumn{8}{c}{Dwarfs} \\
\hline
13976 &  ...  & K2.5V    & \tablefootmark{1}7.95       & 03:00:02.8   & +07:44:59.1 & 42.66$\pm$1.22       & +28.35$\pm$0.18        & 0.83 & 220 \\
16529 &  4  & G5D      &  8.88       & 03:32:50.1   & +23:41:31.9 & 22.78$\pm$1.26       & +32.72$\pm$0.17        & 0.87 & 210 \\
18946 &  ...  & K5D      & \tablefootmark{2}10.13       & 04:03:39.0   & +19:27:18.0 & 23.07$\pm$2.12       & +36.93$\pm$0.26        & 0.75 & 150 \\
19098 &  ...  & K2D      &  \tablefootmark{2}9.29       & 04:05:39.7   & +17:56:15.7 & 19.81$\pm$1.39       & +37.61$\pm$0.05        & 0.88 & 160\\
19148 &  10 & G0V      &  7.85       & 04:06:16.1   & +15:41:53.2 & 21.41$\pm$1.47       & +38.04$\pm$0.17        & 1.08 & 310 \\
19781 &  17 & G5V      &  8.46       & 04:14:25.6   & +14:37:30.1 & 21.91$\pm$1.27       & +39.24$\pm$0.06        & 0.97 & 290 \\
19793 &  15 & G3V      &  8.09       & 04:14:32.3   & +23:34:29.8 & 21.69$\pm$1.14       & +38.21$\pm$0.23        & 1.01 & 320 \\
19934 &  21 & G5D      &  9.15       & 04:16:33.5   & +21:54:26.9 & 19.48$\pm$1.17       & +38.46$\pm$0.19        & 0.92 & 230\\
20130 &  26 & G9V      &  8.63       & 04:18:57.9   & +19:54:24.1 & 23.53$\pm$1.25       & +39.58$\pm$0.06        & 0.93 & 330 \\
20146 &  27 & G8V      &  8.46       & 04:19:08.0   & +17:31:29.1 & 21.24$\pm$1.32       & +38.80$\pm$0.08        & 0.94 & 300 \\
20899 &  73 & G2V      &  7.85       & 04:28:48.3   & +17:17:07.7 & 21.09$\pm$1.08       & +39.37$\pm$0.06        & 1.06 & 470 \\
21112 &  88 & F9V      &  7.78       & 04:31:29.3   & +13:54:12.5 & 19.46$\pm$1.02       & +40.98$\pm$0.31        & 1.13 & 320 \\
22422 & 118 & F8D      &  7.74       & 04:49:32.1   & +15:53:19.5 & 19.68$\pm$0.96       & +42.04$\pm$0.14        & 1.10 & 320 \\
22566 & 143 & F8D      &  7.90       & 04:51:23.2   & +15:26:00.5 & 17.14$\pm$1.00       & +42.92$\pm$0.19        & 1.17 & 250 \\
\bottomrule
\end{tabular}
\tablefoot{Additional sources of V magnitudes:(1) \cite{Koen10}, (2) \cite{Johnson62}.}  
 \end{table*}

\section{Observations and data reduction}\label{sec:data}

All spectra used in this work were downloaded from the  European Southern 
Observatory (ESO) science archive 
facility\footnote{\url{http://archive.eso.org/eso/eso_archive_adp.html}}. The 
giants were observed during ESO program 088.C-0513 with the HARPS 
\citep[High Accuracy Radial velocity Planet Searcher,][]{Mayor03} 
high-resolution spectrograph (R~=~110\,000), situated at the 3.6~m telescope 
in La Silla, Chile. The spectral range covers from 3800 to 6900~\AA, with a 
small gap between 5300--5330~\AA~because of the arrangement of the CCD mosaic. 
The average S/N @6109~\AA~is about 400. All the spectra were reprocessed by the 
last version of the HARPS pipeline (Data Reduction Software version 3.5). We 
only carried out the Doppler correction and the continuum normalization  with 
standard IRAF\footnote{\textit{Image Reduction and Analysis Facility} - IRAF 
provided by \textit{Association of Universities for Research in Astronomy}- 
AURA, EUA.} routines. 

The dwarfs were observed during ESO program 70D-0356 with UVES 
\citep[Ultraviolet and Visual Echelle Spectrograph,][]{Dekker00} at the 8.2m 
Kueyen telescope of the VLT (Very Large Telescope) with and spectral 
resolution of R~$\sim$~60\,000. For consistency, we used spectra acquired 
with the same instrument configuration for all the observations: slit width 
of 0.8$\arcsec$ and central wavelength at 580~nm in the red arm. The 
wavelength coverage is from 4780--6800~\AA with a gap between 
5750--5830~\AA~because of thee arrangement of the CCD mosaic. The S/N varies 
from 150 to 400. Data reduction was carried out with the ESO Reflex environment 
\citep{Freudling13} version 2.3, with the exception of Doppler correction, 
which was performed with IRAF routines. 

We adopt two spectra as solar proxies. For comparison with the giants, we 
use the reflected spectrum of the Jupiter's moon Ganymede\footnote{\url{http://www.eso.org/sci/facilities/lasilla/instruments/harps/inst/monitoring/sun.html}}, 
which was obtained with the same HARPS configuration as the spectra of the 
giants. For the comparison with the dwarfs, we use the UVES solar spectrum\footnote{\url{http://www.eso.org/observing/dfo/quality/UVES/pipeline/solar_spectrum.html}} , which was obtained with the moonlight illuminating the slit. Both 
spectra have, on average, a S/N~$\geq$~300.


\section{Analysis}\label{sec:analysis}

The atmospheric parameters (effective temperature, $T_{\rm eff}$, surface 
gravity, $\log~g$, and microturbulence, $\xi$) of the sample stars were 
determined using two different methods. In this way, we can compare the final 
metallicity scale obtained under different assumptions. The first method, 
hereafter M1 and described in Section \ref{sec:m1}, takes constraints into 
account that do not depend on the classical spectroscopic method, but that 
are still fine-tuned using \ion{Fe}{i} and \ion{Fe}{ii} lines. The second method, 
hereafter M2 and described in Section \ref{sec:m2}, is the classical 
spectroscopic analysis where the parameters are determined using the 
\ion{Fe}{i} and \ion{Fe}{ii} ionization and excitation equilibria. In 
addition, in the implementation of each method, we make use of two different 
line lists. We, therefore, derived four sets of atmospheric parameters for 
each star. Before describing the two methods, we present the two line lists 
and discuss the measurement of equivalent widths (EWs).

\subsection{Line lists selection}

The two line lists that we adopted were assembled with two goals. First, 
we aim to minimize non-LTE effects as it will differentially affect stars 
of different gravities and temperatures. Therefore, the first list contain 
a set of lines for which non-LTE effects would be minimized or neglected, at 
least in the metallicity regime of the Hyades. \cite{Mashon11} evaluated the 
non-LTE line formation of the two ions of iron in cool reference stars, some 
of them, with metallicity comparable to the Hyades. In that study, the authors 
concluded that non-LTE effects are virtually negligible for \ion{Fe}{ii} lines 
(i.e., they affect the abundances by less than 0.01~dex) and are very small for 
\ion{Fe}{i} lines, for stars of metallicity slightly higher than the Sun. We 
selected a total of 42 \ion{Fe}{i} and 15 \ion{Fe}{ii}, from their line list, 
among weak to moderately strong transitions, which were well isolated and as free 
as possible from blending features. We excluded 14 lines of the original list 
for which the measured EW in the Sun was in the saturated regime of the curve 
of growth. \cite{Mashon11} claimed that the accuracy of their iron abundances 
might be affected by the uncertainties in the \emph{gf} values that were used. 
Their \emph{gf} values were obtained from experimental measurements collected 
from different papers. We  improved some of the \emph{gf} values in this 
line list with more recent determinations that were kindly provided by Maria 
Bergemann (private communication). Table~\ref{tab:mash} lists the selected 
lines, their atomic data and, in addition, the equivalent widths and  
individual abundances obtained for the solar reflected spectrum of 
Ganymede using M1. This line list is hereafter referred to as MASH.

The second list was chosen to be suitable for the analysis of giants. Often, 
line lists assembled for the analysis of the Sun may not be optimized for 
giants because of the more pronounced spectral transitions present in these stars. 
We have used a line list with transitions carefully chosen to avoid blends in 
giants and with accurate \emph{gf} values determination provided through the courtesy 
of Dr. Martin Asplund (private communication). This list includes a total of 
34 \ion{Fe}{i} and 7 \ion{Fe}{ii} transitions and hereafter is referred to as 
ASPL. Table~\ref{tab:aspl} lists the lines, as well their atomic data, and the 
equivalent widths and  individual abundances for the solar reflected 
spectrum of Ganymede using M1. There are   15 lines in common between the MASH 
and ASPL lists. Nevertheless, the abundances in the Sun are sometimes distinct 
because of the different atomic data adopted in each list. This difference is 
greater than 0.05~dex for about 20$\%$ of the lines, while for the remaining 
it is about $\sim$~0.02~dex on average.

\subsection{Equivalent width measurements}\label{sec:ews}

We used the code ARES \citep[Automatic Routine for line Equivalent widths in 
stellar Spectra;][]{Sousa07} to perform automatic measurements of the EWs of 
the \ion{Fe}{i} and \ion{Fe}{ii} lines. This code applies a Gaussian fit to 
the profile of the absorption lines, taking  a local continuum 
into account, which is determined on the basis of a set of input parameters provided by the 
user. We tested different combinations of input parameters for both UVES and 
HARPS spectra, and chose those which visually produced the best fits to the 
line profile. The best fits were determined by a visual check and by comparing 
the EWs computed with ARES and the EWs computed with IRAF for a given set of 
stars. These were the same stars selected for the EWs comparison between IRAF 
and ARES (see below). With this approach, the input fitting parameters for ARES were 
optimized. The spectral resolution we use  is such that, in general, 
the instrumental profile dominates the observed profile, and therefore, a Gaussian 
fit can well reproduce the observed profile of the lines. Additionally, we removed 
 all EWs~$<$~5~m\AA\ from our analysis to avoid lines severely affected by noise 
or uncertainties related to the continuum fit, therefore, with larger relative 
errors in the EW measurement. We also removed EWs~$>$~120~m\AA, to avoid the flat 
part of the curve of growth, in which a Gaussian fit may not adequately reproduce 
the observed line profiles. 

The quality of the automatic measurements is comparable to the manual method 
obtained with the task \emph{splot} of IRAF for the majority of the HARPS and 
UVES spectra as shown by \cite{Sousa07}. Notwithstanding, we repeated this 
comparison with the line lists used in our work to verify if the manual EWs 
obtained with \emph{splot} are comparable with those computed using ARES 
for both giants and dwarfs. 

\begin{figure*}
\centering
\includegraphics[height=8cm, width=6cm]{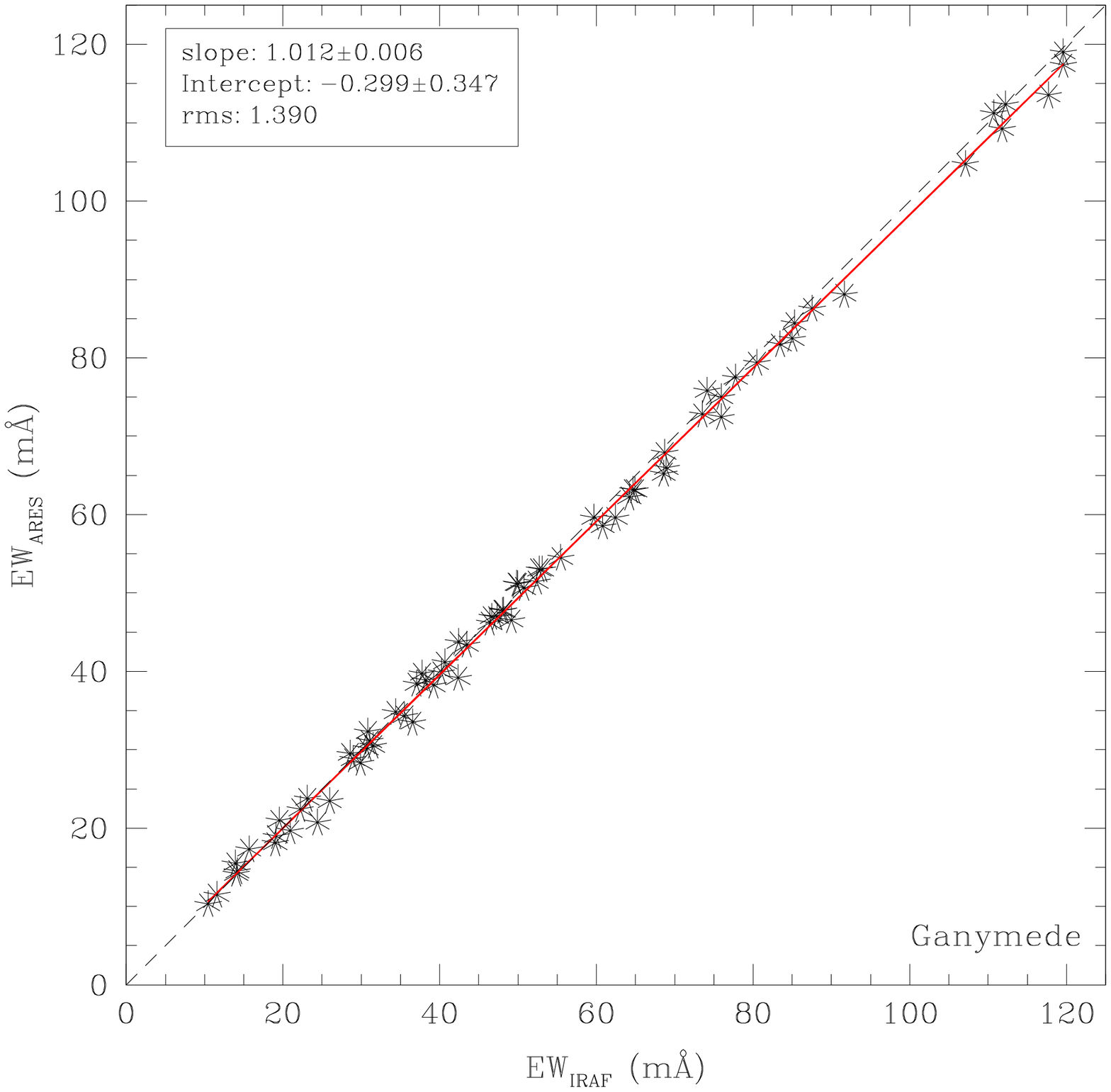}
\includegraphics[height=8cm, width=6cm]{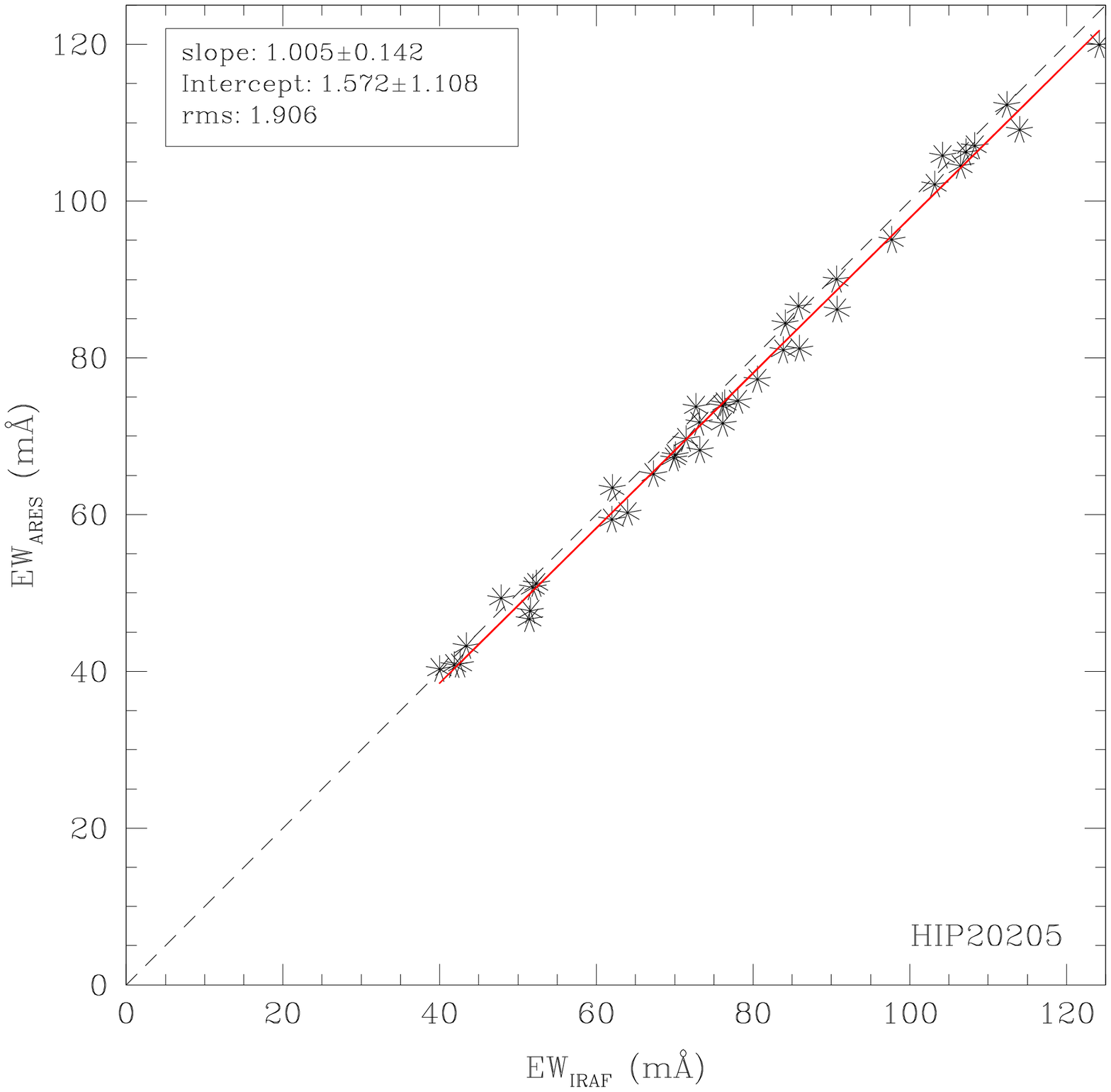}
\includegraphics[height=8cm, width=6cm]{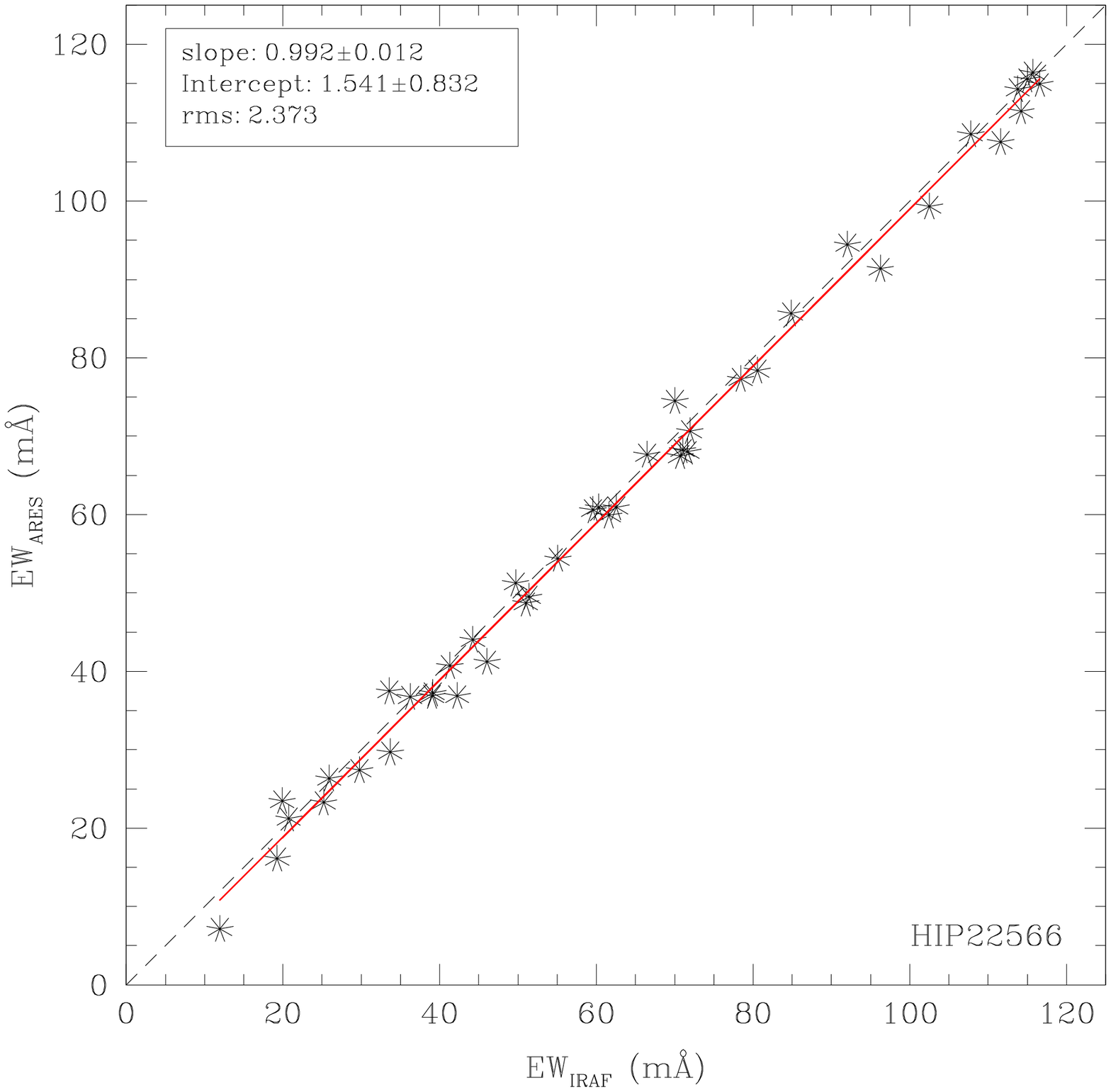}
 \caption{Comparison of EWs obtained with ARES and \emph{splot}. \textbf{Left:} the solar reflected spectrum of Ganymede; \textbf{middle:} the star HIP~20205; and \textbf{right:} the star HIP~22566. The one-to-one relation is shown with the dashed line, and the red line represents the linear regression fit in each plot. The slope, the intercept, and the rms of the fits are also given in each plot.}\label{fig:ews} 
\end{figure*}

Figure \ref{fig:ews} shows the results of such comparison. In addition to the 
solar spectrum, we select as examples the hot dwarf HIP~22566, where the 
continuum placement is more challenging because of the enhanced rotation, and 
the giant HIP~20205, where we have fewer good lines available for the analysis. 
There is in general a good agreement between the automatic and the manual 
measurements. The mean difference between the measurements is 
0.717~$\pm$~1.505~m\AA~for Ganymede, 1.907~$\pm$~1.933~m\AA~for HIP~20205, and 
1.074~$\pm$~2.413~m\AA~for HIP~22566. As seen in Fig.~\ref{fig:ews}, the slope 
and the intercept values of the linear regressions between the two 
sets of EWs are close to one and zero, respectively. 

Although the difference between the two sets of EWs is small, it is not 
negligible, but the comparison indicates the same trend for all stars tested, 
i.e, the EWs obtained with \emph{splot} are systematically higher than those obtained with ARES. To quantify the effect of the EW measurement 
differences on the final metallicities, we computed the abundances for the 
stars shown in Fig~\ref{fig:ews} using both the automatic and the manual 
measurements. The mean difference of the metallicities is $\sim$~0.03~dex for 
the Sun, and $\sim$~0.04~dex for both HIP~22205 and HIP~22566. Based on the 
small differential variation found in the derived abundances, we therefore decided to 
measure all the EWs using ARES. We visually inspected the fits for each line, 
and removed from our analysis those fits that  were judged of poor quality. 
Bad line fits were mainly attributed to poor estimates of the local continuum 
regions or lines that were significantly affected by noise.

\subsection{Method M1: Well-constrained parameters independent of spectroscopy}\label{sec:m1}

The first method used to constrain the atmospheric parameters makes use of 
input values that have been determined independent of the classical spectroscopic method. With this 
new approach, we expect to be able to investigate which are the possible systematics 
that can arise from the classical spectroscopic analysis. 

For the red giants, we adopted the direct determinations of $T_\mathrm{eff}$ by 
\cite{Boyajian09}. Angular diameters were obtained with long-baseline optical 
interferometry and transformed into linear radii using the Hipparcos 
parallaxes. The direct estimations of $T_\mathrm{eff}$ are calculated combining 
these radii and the bolometric flux of the star \citep[see][for the 
definition of direct measurements of $T_\mathrm{eff}$ for FGK stars]{BohmVit81}. 
The bolometric fluxes were determined using the bolometric corrections from 
\cite{AllendePrieto99} and assuming $M_{\rm BOL,\odot}\,=\,$4.74. Effective 
temperatures derived by this method can reach an accuracy of 1$\%$ and are the 
state of the art in $T_\mathrm{eff}$ determinations. 

For the dwarfs, as interferometric angular diameters are not available, we 
obtained $T_\mathrm{eff}$ from photometric calibrations derived with the 
InfraRed Flux Method (IRFM). Although the IRFM provides semidirect 
estimations of $T_\mathrm{eff}$, the temperature scale is almost model 
independent. We use the $JHK$s magnitudes from 2MASS \citep{2MASS} and the 
$(B-V)$ color from the Hipparcos catalog \citep{vanLeeuwen07}. We adopt the 
calibrations from \cite{Casagrande10} to derive photometric temperatures using 
the colors $(V-J)$, $(V-H)$, $(V-K_{s}),$ and $(B-V)$. The adopted photometric 
$T_\mathrm{eff}$ was calculated with an average of the four different 
temperatures estimates weighted by the errors of the calibrations. Effective 
temperatures derived by this method reach and accuracy of a few percent 
\citep{Casagrande14}.

Since the $T_\mathrm{eff}$ of the giants were derived from interferometric measurements 
and those of the dwarfs were derived from IRFM, it is interesting to evaluate how well 
the two methods agree for the giants. The IRFM $T_\mathrm{eff}$ for the giants are shown 
in Table~\ref{teffphot}. \cite{daSilva06} showed that 2MASS colors are unsuitable to 
determine $T_\mathrm{eff}$ of bright stars. For this reason, we used only the $(B-V)$ 
color combined with the calibrations of \cite{GonzalezHer09}, \cite{Ramirez05} and \cite{Alonso96}, 
which are more appropriated for evolved stars. Interferometric effective temperatures 
($T_\mathrm{eff}$(int)) obtained from \cite{Boyajian09} are also shown for comparison in 
Table~\ref{teffphot}. The average IRFM $T_\mathrm{eff}$ are in a excellent agreement with 
the interferometric values within less than 50~K, as shown in the last column of 
Table~\ref{teffphot}. A difference between individual IRFM and interferometric $T_\mathrm{eff}$ 
as large as 100~K is found only for the star HIP~20889 when using the calibration of 
\cite{GonzalezHer09}, which gives values systematically lower than the others.

\begin{table*}[!htbp]
\ra{1.2}
\small
 \caption{IRFM $T_\mathrm{eff}$ obtained with three different calibrations for the giants of the Hyades. $T_\mathrm{eff}$(int) obtained from \cite{Boyajian09} are also shown for comparison. The $<T_\mathrm{eff}>$ corresponds to the average of the three IRFM $T_\mathrm{eff}$. The last column corresponds to the difference between the $<T_\mathrm{eff}>$ and $T_\mathrm{eff}$(int).} 
 \label{teffphot}
\centering
\begin{tabular}{cccccccccc}
\toprule
HIP   & $T_\mathrm{eff}$(GH09) & $T_\mathrm{eff}$(AL96) & $T_\mathrm{eff}$(RM05) & $<T_\mathrm{eff}>$ & $T_\mathrm{eff}$(int) & $\Delta$$T_\mathrm{eff}$  \\
\midrule
20205 & 4782            & 4874           & 4892            & 4849           & 4844     & 5              \\
20455 & 4778            & 4870           & 4888            & 4845           & 4826     & 19             \\
20889 & 4720            & 4810           & 4820            & 4783           & 4827     & -44            \\
\bottomrule
\end{tabular}
\tablefoot{GH09, AL96 ,and RM05 stands for the calibrations of \cite{GonzalezHer09}, \cite{Alonso96} and \cite{Alonso96}, respectively.}
\end{table*}

Surface gravities were determined according to the following equation:
\begin{equation}
\rm{log}(g_{*}/g_{\odot})=\rm{log}(M_{*}/M_{\odot}) +4~\rm{log}(T_{\rm{eff*}}/T_{\rm{eff\odot}})-\rm{logg}(L_{*}/L{\odot}).
\end{equation}
For the dwarfs, the masses were computed with the theoretical evolutionary 
tracks of \cite{Girardi2000} and a Bayesian estimation method, which takes  the error of each quantity for the mass determination into 
account \citep[see][for 
details]{PARAM}. Luminosities were taken from \cite{Bruijne01} and the 
$T_\mathrm{eff}$ are those computed as mentioned above. For the giants we 
adopted a value of mass of 2.48~$M_\sun$; the mass of a clump giant in a 
\citet{Girardi02} isochrone of 625 Myr and [Fe/H] = +0.13 \citep[see][for 
 details]{Smiljanic12}. 

One free parameter that can not be constrained by our observational knowledge 
of the cluster is the microturbulence velocity ($\xi$). For M1, we fixed $\xi$ 
adopting predictions obtained with 3D atmospheric models. The details on how 
these values were obtained are described in Section \ref{sec:micro}.  

To calculate the metallicities, we used 1D-LTE plane-parallel atmospheric 
models computed using the Linux version of the ATLAS9 code 
\citep{Sbordone04,Sbordone05} originally developed by Kurucz 
\citep[see, e.g.,][]{Kurucz93} and adopting the ODFNEW opacity distribution 
from \cite{CastelliKurucz03}. The mixing length parameter adopted was 1.25 and 
no overshooting was considered for both giants and dwarfs. Chemical abundances 
of \ion{Fe}{i} and \ion{Fe}{ii} were derived using WIDTH package 
\cite{Kurucz93}, under some minor optimizations to facilitate handling the 
input data. 

Our main innovation in method M1 is that the spectroscopy independent 
parameters described above were used as input values and subjected to a 
further controlled fine-tuning. The best set of parameters for each star was 
determined as follows. First, we allowed the stellar parameters to vary within 
a range of conservative errors to find the best agreement between the 
abundances of the \ion{Fe}{i} and \ion{Fe}{ii} lines. The ranges were 
$\pm$50~K for $T_\mathrm{eff}$, with steps of 10~$\mathrm{K}$; 
$\pm$0.10~$\mathrm{dex}$ for $\log~g$, with steps of 0.05~$\mathrm{dex}$; and 
$\pm$0.10~$\mathrm{km~s^{-1}}$ for $\xi$, with steps of 
0.05~$\mathrm{km~s^{-1}}$. These ranges were chosen because they represent 
typical error values of the classical abundance analysis. We tested variations 
using smaller steps, but no major improvement on the final solution was found. 
Next, after reaching a solution, we applied a 2$\sigma$-clipping to remove 
lines classified as outliers with respect to the average abundances of the 
previous iteration. The final set of atmospheric parameters is given by 
looking for the best possible agreement between the abundances of 
\ion{Fe}{i} and \ion{Fe}{ii} in the 3D-plane 
$T_\mathrm{eff}$ -- $\log~g$ -- $\xi$. The stellar parameters and 
metallicities derived with this approach, for giants and dwarfs, with  both MASH 
and ASPL lists, are shown in Table~\ref{tab:paramm1}.  

\begin{sidewaystable*}
\caption{Stellar parameters and iron abundances for giants and dwarfs derived according to method M1. The left side of the table shows the results obtained with the MASH line list and the right side the results obtained with the ASPL line list.}\label{tab:paramm1}
\centering
\ra{1.1}
\small
\begin{tabular}{@{}c c c c c c c c   c c c c c c c c c c c c c c c c c c c c c c@{}}\toprule[1.0pt]
&  \multicolumn{7}{c}{\bf{MASH}}  &  \multicolumn{7}{c}{\bf{ASPL}} \\
\midrule
HIP & $T_\mathrm{eff}$ &  $\log~g$ &  $\xi$ & log$\epsilon_{\mathrm{FeI}}$  & $\mathrm{N(FeI)}$ &  log$\epsilon_{\rm{FeII}}$   & $\mathrm{N(FeII)}$ & $T_\mathrm{eff}$ &  $\log~g$ &  $\xi$ & log$\epsilon_{\mathrm{FeI}}$  & $\mathrm{N(FeI)}$ &  log$\epsilon_{\rm{FeII}}$   & $\mathrm{N(FeII)}$ \\
\cmidrule(r){2-8}\cmidrule(l){9-15}
20205        & 4874 & 2.61 & 1.30 & 7.62$\pm$0.10 & 22 & 7.62$\pm$0.09 & 11  & 4884 & 2.61 & 1.30 & 7.59$\pm$0.06 & 20 & 7.59$\pm$0.06 & 7 \\
20455        & 4876 & 2.59 & 1.30 & 7.59$\pm$0.10 & 22 & 7.62$\pm$0.09 & 11  & 4876 & 2.59 & 1.35 & 7.56$\pm$0.07 & 20 & 7.56$\pm$0.06 & 7 \\ 
20889        & 4817 & 2.65 & 1.35 & 7.70$\pm$0.09 & 18 & 7.70$\pm$0.10 & 10  & 4827 &  2.60 & 1.35 & 7.63$\pm$0.08 & 18 & 7.63$\pm$0.06 & 6 \\ 
\hline
\noalign{\smallskip}
\it{average}  &                 &      &      & \it\textbf{7.64$\pm$0.06} &&  \it\textbf{7.65$\pm$0.05} &        &      &      &      & \it\textbf{7.59$\pm$0.04} & &   \it\textbf{7.59$\pm$0.04}   \\
\hline
\midrule
HIP & $T_\mathrm{eff}$ &  $\log~g$ &  $\xi$ & log$\epsilon_{\mathrm{FeI}}$  & $\mathrm{N(FeI)}$ &  log$\epsilon_{\rm{FeII}}$   & $\mathrm{N(FeII)}$ & $T_\mathrm{eff}$ &  $\log~g$ &  $\xi$ & log$\epsilon_{\mathrm{FeI}}$  & $\mathrm{N(FeI)}$ &  log$\epsilon_{\rm{FeII}}$   & $\mathrm{N(FeII)}$ \\
\cmidrule(r){2-8}\cmidrule(l){9-15}
18946 & 4661  &  4.54 & 0.75 & 7.63$\pm$0.09   & 21 & 7.63$\pm$0.11 & 6 & 4691  & 4.64 & 0.70 & 7.56$\pm$0.09   & 17 & 7.56$\pm$0.21 & 5  \\
13976 & 5023  &  4.51 & 0.85 & 7.69$\pm$0.07   & 22 & 7.69$\pm$0.11 & 8 & 5013  & 4.61 & 0.80 & 7.64$\pm$0.08   & 22 & 7.64$\pm$0.08 & 5  \\
19098 & 5178  &  4.59 & 0.85 & 7.62$\pm$0.09   & 22 & 7.62$\pm$0.06 & 9 & 5138  & 4.54 & 0.80 & 7.63$\pm$0.09   & 24 & 7.63$\pm$0.06 & 7  \\
16529 & 5237  &  4.51 & 0.80 & 7.64$\pm$0.10   & 22 & 7.64$\pm$0.08 & 10 & 5207  & 4.51 & 0.85 & 7.62$\pm$0.08   & 22 & 7.62$\pm$0.05 & 6  \\
19934 & 5361  &  4.57 & 0.90 & 7.64$\pm$0.06   & 23 & 7.64$\pm$0.06 & 9  & 5341  & 4.57 & 0.85 & 7.62$\pm$0.08   & 27 & 7.62$\pm$0.04 & 6  \\
20130 & 5531  &  4.45 & 0.95 & 7.66$\pm$0.09   & 28 & 7.66$\pm$0.07 & 10 & 5511  & 4.55 & 0.90 & 7.62$\pm$0.06   & 26 & 7.62$\pm$0.07 & 7  \\
20146 & 5563  &  4.33 & 1.00 & 7.67$\pm$0.08   & 26 & 7.67$\pm$0.08 & 10 & 5553  & 4.43 & 0.95 & 7.62$\pm$0.07   & 28 & 7.62$\pm$0.04 & 6  \\
19781 & 5641  &  4.35 & 0.85 & 7.69$\pm$0.09   & 30 & 7.69$\pm$0.08 & 10 & 5621  & 4.40 & 0.90 & 7.60$\pm$0.06   & 27 & 7.60$\pm$0.05 & 7  \\
19793 & 5831  &  4.31 & 1.00 & 7.73$\pm$0.08   & 29 & 7.73$\pm$0.09 & 10 & 5781  & 4.41 & 0.95 & 7.64$\pm$0.08   & 28 & 7.64$\pm$0.05 & 7  \\
20899 & 5916  &  4.31 & 0.90 & 7.66$\pm$0.06   & 26 & 7.66$\pm$0.08 & 10 & 5886  & 4.31 & 0.95 & 7.61$\pm$0.08   & 27 & 7.61$\pm$0.04 & 8 \\
19148 & 6021  &  4.37 & 1.00 & 7.61$\pm$0.06   & 28 & 7.61$\pm$0.09 & 10 & 5961  & 4.37 & 0.85 & 7.62$\pm$0.08   & 28 & 7.62$\pm$0.04 & 7 \\
22422 & 6074  &  4.34 & 1.05 & 7.66$\pm$0.09   & 32 & 7.66$\pm$0.08 & 10 & 6004  & 4.34 & 1.00 & 7.63$\pm$0.06   & 30 & 7.63$\pm$0.01 & 6 \\
21112 & 6161  &  4.36 & 1.10 & 7.60$\pm$0.09   & 34 & 7.63$\pm$0.04 & 8  & 6161  & 4.26 & 1.05 & 7.60$\pm$0.06   & 29 & 7.60$\pm$0.03 & 7 \\
22566 & 6251  &  4.30 & 1.10 & 7.69$\pm$0.08   & 27 & 7.69$\pm$0.09 & 9  & 6211  & 4.30 & 1.10 & 7.66$\pm$0.08   & 26 & 7.66$\pm$0.07 & 6 \\
\hline
\noalign{\smallskip}
\it{average}  &                 &      &      & \it\textbf{7.66$\pm$0.04} & &   \it\textbf{7.66$\pm$0.03} &        &      &      &      & \it\textbf{7.62$\pm$0.02} & &   \it\textbf{7.62$\pm$0.02}   \\
\bottomrule
\end{tabular}
\end{sidewaystable*}

The main aspect of M1 is that the final parameters shown in Table~\ref{tab:paramm1} can 
vary within a very narrow range, constrained by independent methods. Thus, the errors for 
the stellar parameters in M1 are the uncertainties in the input parameters. An error of about 
$\sim$50~K corresponds to the error of the IRFM calibrations and is comparable with the 
interferometric errors presented in the work of \cite[][see their Table~4]{Boyajian09}. For 
$\log~g$, we adopted an error of 0.10~dex because we have very small errors in the parallaxes, 
thus the main source of error in the gravities comes from our mass determination. An error of 
20$\%$ in the masses  changes the surface gravities by $\sim$~0.10~dex. Since the evolutionary 
status of the Hyades is well known, we consider the gravity error of 0.1~dex as conservative and 
most likely the highest source of uncertainty is in the mass-loss estimate for the giants. For 
$\xi$ we adopted an error of 0.10~$\mathrm{km~s^{-1}}$, which corresponds to the  uncertainty 
estimated with the 3D microturbulence calibration (see Section~\ref{sec:micro}). The dispersion 
for the \ion{Fe}{i} and \ion{Fe}{II} abundances shown in Table~\ref{tab:paramm1} correspond to 
the standard deviation of each abundance distribution.

We also evaluate how the uncertainty in the physical parameters of the stars affect the retrieved 
abundances when using M1. To this end, we assumed the parameters presented in Table~\ref{tab:paramm1} 
for the giant HIP~20205 and the dwarf HIP~19148,  and then vary $T_\mathrm{eff}$, $\log~g$ and $\xi$ 
separately for both ASPL and MASH lists. The results are given in Table~\ref{tab:errorM1}. We emphasize 
that the variations in \ion{Fe}{i} and \ion{Fe}{ii} follow the same direction for both, the 
giant and  main-sequence stars.

\begin{table*} [!htbp] 
\ra{1.2}
\small
 \caption{Abundance changes ($\Delta$[X/H]) in a giant (HIP~20205) and in a main-sequence (HIP~19148) star in response to variations of $T_\mathrm{eff}$, $\mathrm{log g}$ and $\xi$.}
 \label{tab:errorM1}
\centering
\begin{tabular}{cccccccccc}
\toprule
   \multicolumn{7}{c}{ASPL LIST}\\
\midrule
El.  &  $T_\mathrm{eff}$+50~K  & $T_\mathrm{eff}$-50~K & $\log~g$+0.10~dex   & $\log~g$-0.10~dex & $\xi$+0.10~$\mathrm{km~s^{-1}}$  & $\xi$-0.10~$\mathrm{km~s^{-1}}$   \\
\hline
 \multicolumn{7}{c}{HIP~20205}\\

\ion{Fe}{i}  & 0.03 & -0.01 &   0.01 & 0.00 & -0.04 & 0.06 \\                           
\ion{Fe}{ii} & -0.04 & 0.03 &   0.05 & -0.05 & -0.02 &  0.01 \\
\hline
  \multicolumn{7}{c}{HIP~19148}\\

\ion{Fe}{i}  & 0.04 & -0.04 & -0.01 & 0.01 & -0.02 & 0.06 \\                            
\ion{Fe}{ii} & -0.01 & 0.01 & 0.03 & -0.03 & -0.02 & 0.01  \\
\midrule
   \multicolumn{7}{c}{MASH LIST} \\
\hline
 \multicolumn{7}{c}{HIP~20205}\\

\ion{Fe}{i}  & 0.03 & -0.03 &   0.00 & -0.01 & -0.05 & 0.04 \\                          
\ion{Fe}{ii} & -0.04 & 0.04 & 0.05 & -0.05 & -0.04 & 0.04 \\
\hline
 \multicolumn{7}{c}{HIP~19148}\\

\ion{Fe}{i}  & 0.04 & -0.03 & -0.01 & 0.01 & -0.02 & 0.02\\                             
\ion{Fe}{ii} & -0.01 &  0.02 & 0.04 & -0.03 & -0.02 & 0.02  \\
\bottomrule
\end{tabular}
\end{table*}

We obtain a good agreement between the metallicity of giants and dwarfs with 
this method. Moreover, since we are able to retrieve stellar parameters with 
a good agreement between \ion{Fe}{i} and \ion{Fe}{II} abundances, we do not 
see evidence for significant LTE departures either among the giants or the 
dwarfs according to this method. In Table~\ref{tab:paramm1}, we 
find the metallicities obtained using MASH list are slightly higher 
than those using the ASPL list. The systematic offset of about 
$\sim$~0.04-0.06~dex is likely related to the different selection of lines, 
but can also be considered part of the internal errors of the analysis, as an 
accuracy better than 0.05~dex can hardly ever be achieved without a line-by-line 
differential analysis. We further discuss this behavior in 
Section~\ref{sec:discussion}.

\subsubsection{Probing microturbulence velocities with 3D models}\label{sec:micro}

Aside from the different techniques one can adopt to determine the 
microturbulence velocity, it is important to recall that this parameter arises 
from a limitation of the classical 1D model atmospheres into fully describing 
all the velocity fields present in the stellar photosphere. As a consequence, the 
inclusion of an extra velocity field is required to describe the broadening observed in the lines 
placed in the partly saturated regime of the curve of growth 
\citep{StruveElvey34,vanParadijs72}. In practice, it is essential to use this 
parameter to obtain the same abundance for lines with small and large EWs. 
Thus, the optimal value of $\xi$ is obtained by imposing the absence of a 
correlation between the abundances and the EWs of a set of lines. This analysis, 
of course, depends on a good statistics of weak and moderately strong lines. This 
is usually a challenge in the simultaneous analysis of giant and dwarf stars.

One alternative to overcome the limitations cited above is to use 3D 
atmospheric models to predict $\xi$. Three-dimensional models treat convection 
in a physically consistent way, without the need of defining free parameters 
like $\xi$. This approach has been adopted by \cite{Steffen09,Steffen13}, and 
it is based on the comparison between lines computed with spectral synthesis 
using 3D and 1D models. Given a sample of spectral lines, the EWs computed 
from the 3D model with a fixed abundance are taken to represent the 
observation. For each line, the 1D abundance is obtained by matching the 
``observed'' 3D EW with the synthetic line profiles derived from the 1D models
(using exactly the same atomic line parameters as in the 3D synthesis).
The best value of $\xi$ for use with 1D models is taken to be that which 
eliminates the correlation between line strength and derived 1D abundance.
In fact, the best $\xi$ depends somewhat on the choice of the line list.
Ideally, the lines should be insensitive to temperature fluctuations and
have similar properties as the lines to be used for the abundance 
determinations. High-excitation \ion{Fe}{i} lines and \ion{Fe}{ii} lines
are an obvious choice.

We used 3D hydrodynamic models taken from the CIFIST grid 
\citep{Ludwig09}\footnote{An extended grid with respect to 2009.} computed 
with the $\rm{CO^{5}BOLD}$ code\footnote{COnservative COde for the 
COmputation of COmpressible COnvection in a BOx of L Dimensions, L=2,3. 
\url{http://www.astro.uu.se/~bf/co5bold_main.html}} \citep{Freytag12}. The 
metallicity of all selected models is solar since this grid does not  have 
models available for stars more metal-rich than the Sun yet. Indeed, the differences 
between temperature stratification of 3D and 1D models are expected to be 
small at slightly super solar metallicity. At this regime, the line opacities 
act toward heating the optically thin layers and, as a result, the mean 
temperature of the layer is close to radiative equilibrium. We do not expect 
the small metallicity differences to have any noticeable effect on the derived
microturbulence velocities.

The required 1D hydrostatic reference models were computed with LHD code 
\citep{Caffau&Ludwig07} using the same stellar parameters and opacity scheme 
as the 3D models. Table~\ref{tab:micro} shows the stellar parameters of the 
computed models used for two giants, four subgiants, and six dwarfs, which were 
selected to be representative of the Hyades HR diagram. For giants and dwarfs, 
the models were selected to cover as well as possible the range of 
temperatures and gravities of our sample (e.g., Table~\ref{tab:paramm1}). For 
the subgiants, the models cover the cool part of the temperature range 
expected for subgiants in the Hyades, but have slightly higher $\log~g$ 
($\sim$ 3.50) with respect to what is expected from the cluster theoretical 
isochrone ($\log~g \sim$ 3.0). Models for subgiants with $\log~g$ smaller 
than 3.50 were not available in the grid.

For computing the $\xi$ values, we used 30 \ion{Fe}{i} plus 7 \ion{Fe}{ii} 
lines from the ASPL list and applied a method similar to method~3a of 
\cite{Steffen13}. The selected lines are identified in Table\,\ref{tab:aspl}. Here 
again, lines with EW $>$ 120m\AA~were rejected to make this analysis compatible 
with the clipping criteria that were adopted in both M1 and M2. Briefly, for each 
line, we computed the equivalent width using the 3D model, $W_\mathrm{3D}$. Then, we 
computed for the very same lines a 2D curve of growth from the adopted 1D reference 
model, $W_\mathrm{1D}$($\Delta$log$\epsilon$, $\xi$), where $\Delta$log$\epsilon$ is 
the abundance difference with respect to the original abundance used in the 3D 
spectrum synthesis. This grid allows us to find, by interpolation for given $\xi$ 
value, $\Delta$log$\epsilon_i$ for each line $i$ from the condition 
$W_{\mathrm{3D}_{i}}=W_{\mathrm{1D}_{i}}$. In other words, this abundance correction, 
$\Delta$log$\epsilon_i$, is the difference of the abundance computed in the 1D model 
by fitting the equivalent width of the 3D line, and the true abundance used in the 
3D spectrum synthesis. We computed $\Delta$log$\epsilon_i$ for a grid of microturbulence 
values ranging from 0 to 2~$\mathrm{km~s^{-1}}$ with intervals of 0.1~$\mathrm{km~s^{-1}}$. 
The only exception was the giant with $T_\mathrm{eff}$~=~5000~K, where the grid was 
ranging from 0 to 3~$\mathrm{km~s^{-1}}$ in steps of 0.15~$\mathrm{km~s^{-1}}$. 

\begin{figure*}[htbp]
\centering
\begin{tabular}{cc}
\includegraphics[width=0.50\textwidth]{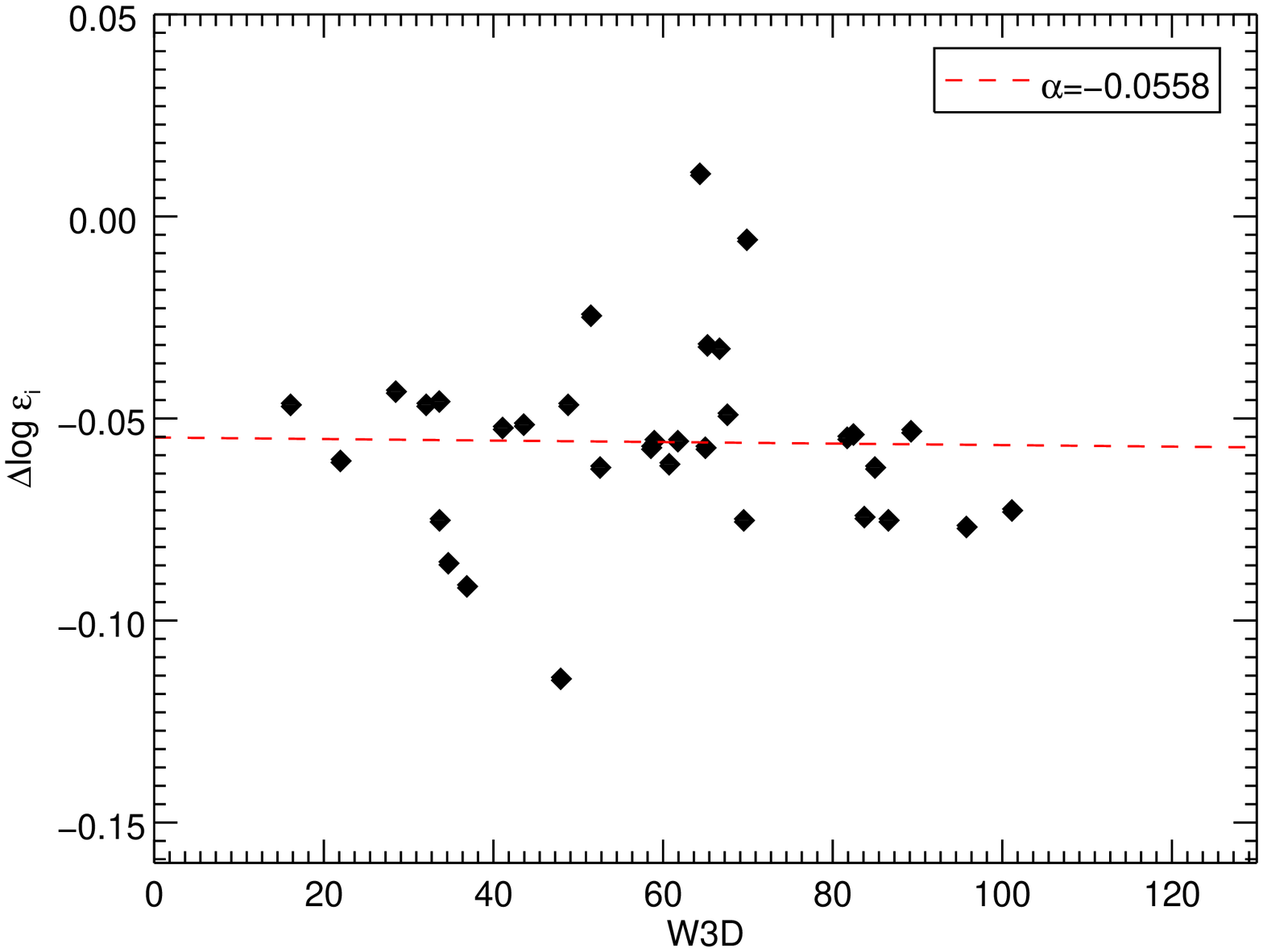} &
\includegraphics[width=0.50\textwidth]{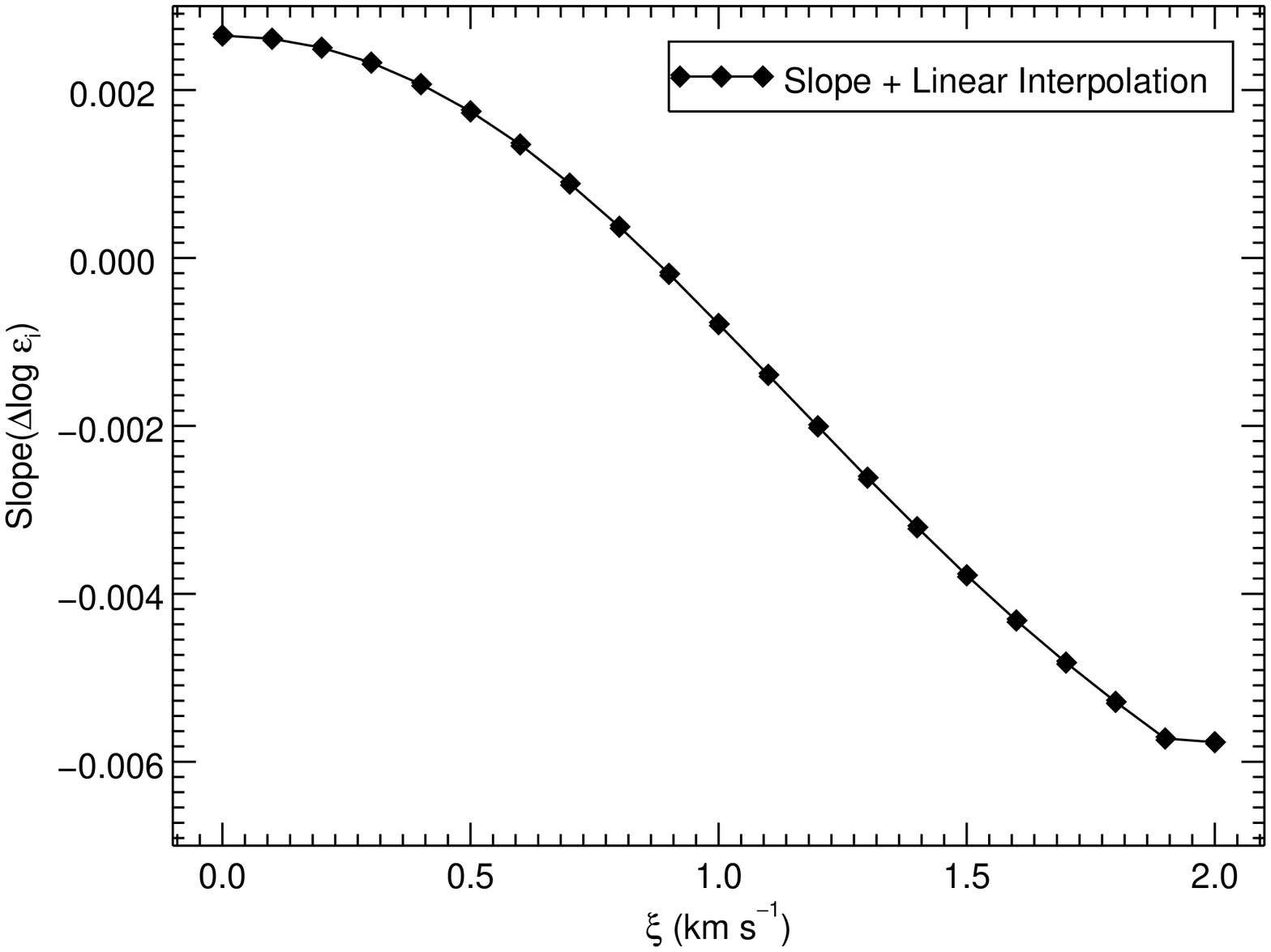} \\
\end{tabular}
 \caption{Microturbulence determination for the model of the dwarf with 
$T_\mathrm{eff}$~=~5000~K and $\log~g$~=~4.5~dex. \textbf{Left:} 
$\Delta$log$\epsilon_i$ versus the 3D equivalent width for 
$\xi$~=~0.90~$\mathrm{km~s^{-1}}$ (symbols). The dashed line represents the 
linear regression to the data points. \textbf{Right:} slope of the linear 
regression as a function of microturbulence. The optimal microturbulence is 
derived by the condition that the slope is zero.}\label{fig:xi}
\end{figure*}

For each $\xi$ value, we plot $\Delta$log$\epsilon_i$ as a function of 
$W_\mathrm{3D}$ (see Fig.~\ref{fig:xi}, left side, for an example). This graph 
illustrates the classical concept of defining the microturbulence. We then 
determine the slope of the linear regression from each of 
the plots described above. These slopes are plotted against the 
corresponding microturbulence values, as shown in Fig.~\ref{fig:xi} 
(right side). The best value of $\xi$ for each star is taken to be 
that where the slope is zero in this curve. By following the procedure 
described here, we estimated the microturbulence values for each of 
the model stars with parameters as listed in Table~\ref{tab:micro}. 

\begin{table} [htbp]
\ra{1.1}
\small
 \caption[Micro Results.]{Microturbulence values computed from 3D models in correspondence with 1D models.}\label{tab:micro}
\begin{center} 
\begin{tabular}{c c c c }\toprule
Star & $T_\mathrm{eff}$ K &  $\log g$ dex & $ \left \langle \xi \right \rangle$ \\
\midrule
Giant    & 4477 & 2.5 & 0.89 \\
Giant    & 4968 & 2.5 & 1.40 \\
Subgiant & 4582 & 3.2 & 0.84 \\
Subgiant & 4923 & 3.5 & 0.90 \\
Subgiant & 5432 & 3.5 & 1.13 \\
Subgiant & 5884 & 3.5 & 1.27 \\
Dwarf    & 4509 & 4.5 & 0.71 \\
Dwarf    & 4982 & 4.5 & 0.85 \\
Dwarf    & 5488 & 4.5 & 0.87 \\
Dwarf    & 5865 & 4.5 & 0.95 \\ 
Dwarf    & 6233 & 4.5 & 1.02 \\  
Dwarf    & 6456 & 4.5 & 1.10 \\ 
\bottomrule
\end{tabular}
\end{center}
 \end{table}

We used the data listed in Table \ref{tab:micro} to establish an empirical 
relation of the microturbulence as a function of effective temperature and 
surface gravity. We tested different functional forms for the calibration 
and adopted that which gave the best statistical response. We searched for 
a calibration that shows no significant trend in the residuals distribution, 
but we have also assessed the quality of the fit by inspecting its correlation 
coefficient ($R$), the standard deviation $\sigma$ of the fit, and the 
$p-$values for each term of the calibration. The best fit was found for the 
following equation:

\begin{equation}
\begin{split}
\xi \,\,(\rm{km\,s^{-1}}) = 0.998 + 3.16\times10^{-4}\,X - 0.253\,Y\\ -2.86 \times 10^{-4}\,X\,Y + 0.165\,Y^2\, ,
\label{eq:micro}
\end{split}
\end{equation}
where $X \equiv T_{\rm eff} - 5500$ [K] and $Y \equiv \log g - 4.0$.
The rms scatter of the residuals of this relation is 
0.05~$\mathrm{km~s^{-1}}$. However, we assume as the total uncertainty of the 
calibration a more conservative value of 0.10~$\mathrm{km~s^{-1}}$ because of the 
small number of stars used in the fit. Although other calibrations in the 
literature \citep[e.g.,][]{E93,F98,Bruntt12} have considered a larger number 
of stars, their estimates of microturbulence are based on spectroscopic 
analyses with different line lists, while Eq.~(\ref{eq:micro}) presents a 
relation that reflects predictions from 3D models with a single line list,
albeit for a limited set of stellar parameters.  

The results for the microturbulence values on Table~\ref{tab:micro} are 
consistent with those presented in Table~3 of \cite{Steffen13}, although these 
authors adopted a different selection of lines (\ion{Fe}{i} lines with a 
lower excitation potential greater than 2\,eV only) and a slightly different 
method (their method~3b). In this case, the microturbulence is given by the 
value that minimizes the scatter of the abundance corrections 
$\Delta$log$\epsilon_i$. \cite{Steffen13} prefer method~3b over 3a, arguing 
that 3a is more susceptible to details of the spectral line sample. Nevertheless, 
we decided to follow 3a because it is closer to the usual analysis applied to 
determine $\xi$ in the literature, and thus facilitates the comparison with 
values derived in a classical spectroscopic analysis. 

We remark that our values display a trend for increasing $\xi$ toward higher 
temperatures and (perhaps) lower gravities. However, the limited number of 
stars tested here is not sufficient to provide more than a rough idea of the 
microturbulence behavior across larger areas of the HR diagram. More details 
about these trends can be found in \cite{Steffen13}.

\subsection{Method M2: The classical spectroscopic analysis}\label{sec:m2}

The second method used to constrain the atmospheric parameters is the 
classical spectroscopic analysis based on the ionization and excitation 
equilibria of the \ion{Fe}{i} and \ion{Fe}{ii} lines. We want to investigate 
whether a consistent metallicity scale between giants and dwarfs can be 
recovered through this method. Effective temperatures are calculated, forcing 
the \ion{Fe}{i} line abundances to be independent of the excitation potential, 
i.e., forcing the excitation equilibrium. The microturbulence velocity is 
determined, forcing the \ion{Fe}{i} abundances to be independent of the EWs. 
Surface gravities are calculated, forcing the lines of \ion{Fe}{i} and 
\ion{Fe}{ii} to produce the same abundance, fulfilling the ionization 
equilibrium. As a consequence, the metallicity ([Fe/H]) is obtained as a 
spin-off of this procedure. All these criteria must converge in a fully 
consistent way and the final solution should be independent of the initial 
input parameters and iteration path. 

\begin{figure*}[htbp]
\centering
\begin{tabular}{cc}
\includegraphics[width=0.50\textwidth]{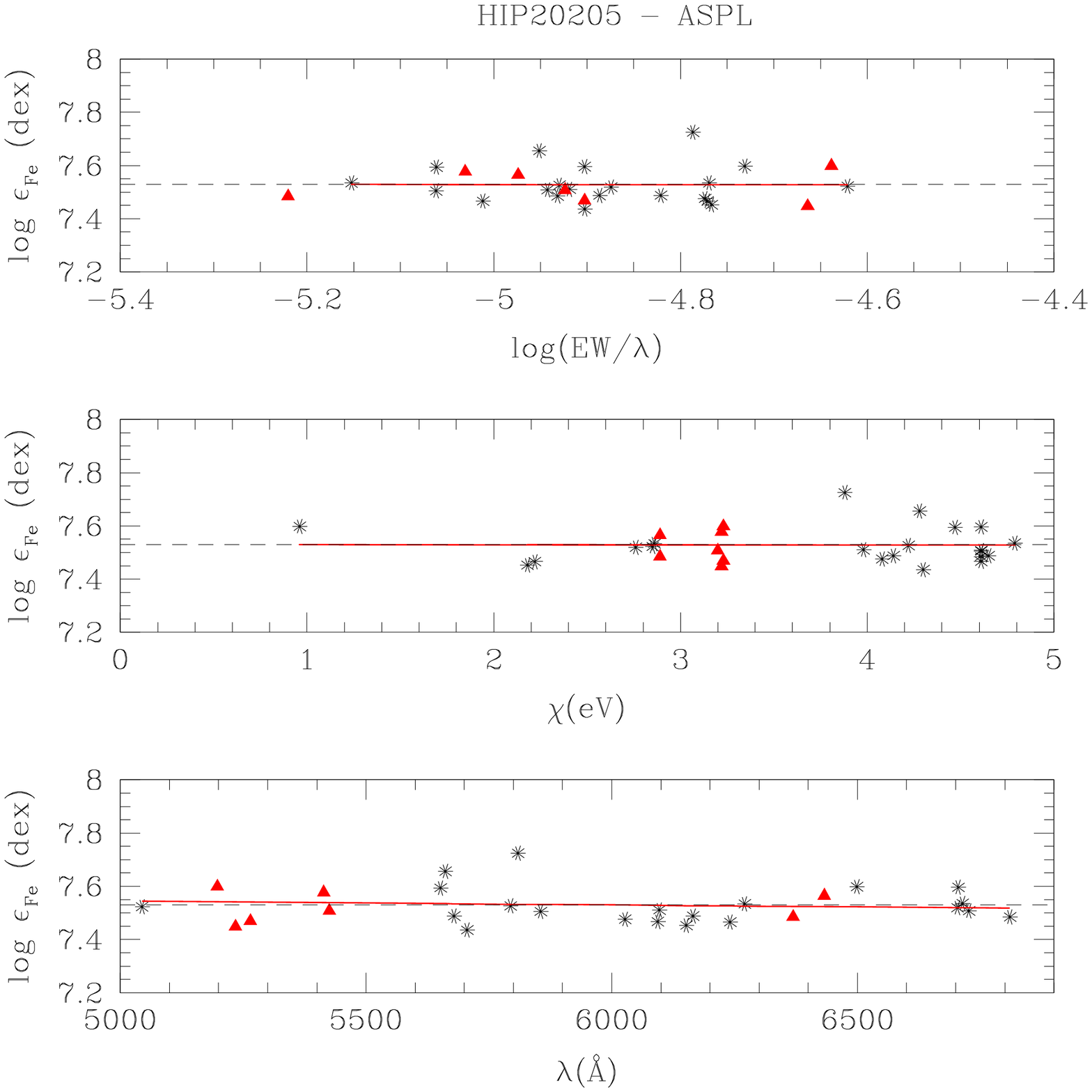} &
\includegraphics[width=0.50\textwidth]{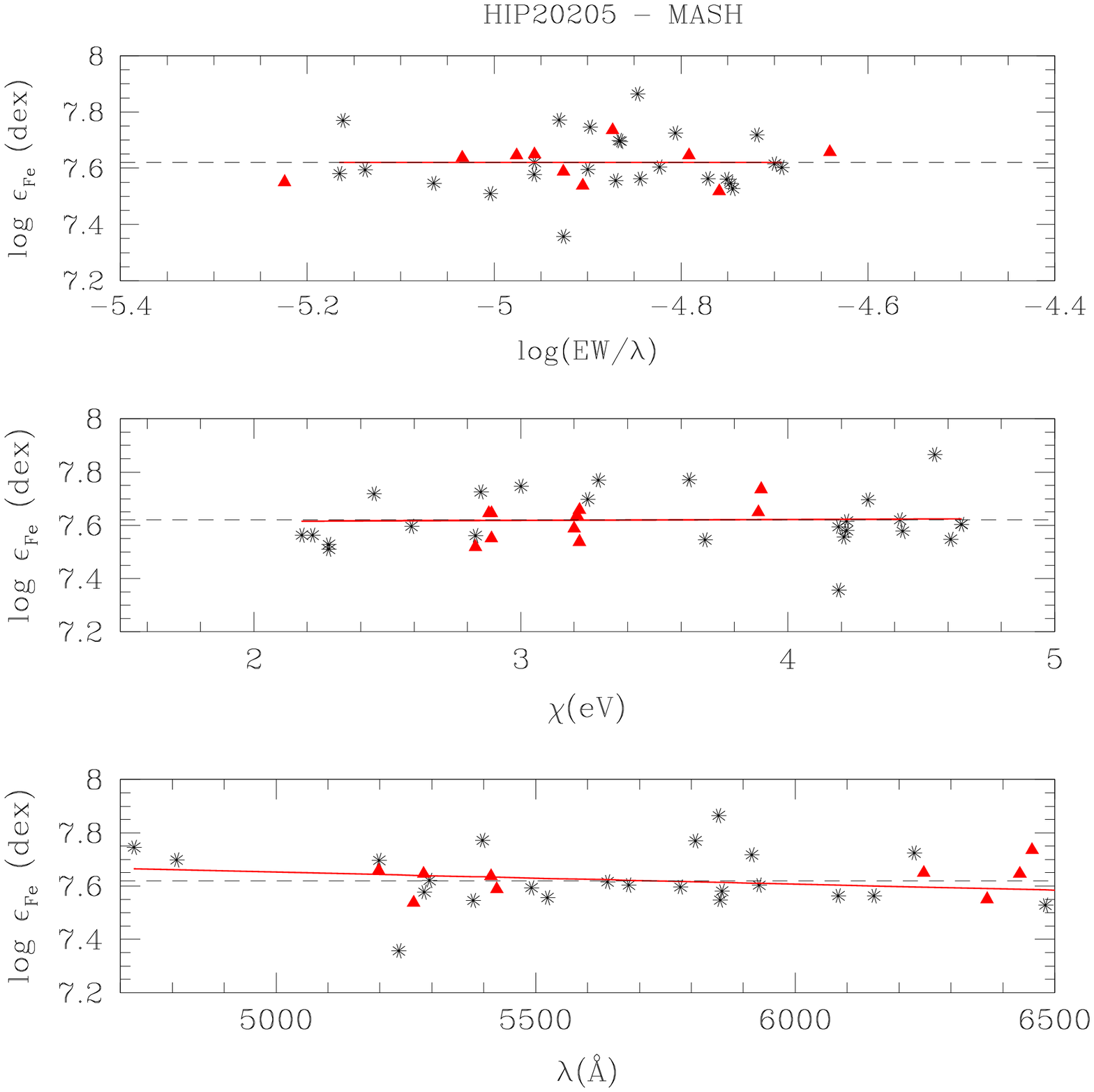} \\
\end{tabular}
\caption{ \ion{Fe}{i} and \ion{Fe}{ii} abundances as a function of the excitation potential ($\chi$), the logarithm of the reduced EW and the wavelength. The continuous red line is the regression between the quantities shown in each plot. The dashed line corresponds to the \ion{Fe}{II} final abundance derived according to M2. In all plots, the black asterisks correspond to the \ion{Fe}{i} lines and the red filled triangles correspond to the \ion{Fe}{II} lines. \textbf{Left:} spectroscopic analysis for HIP~20205 using ASPL list. \textbf{Right:} spectroscopic analysis for HIP~20205 using MASH list.}\label{fig:m2}
\end{figure*}

For this method, we use the same models as in Section~\ref{sec:m1}, i.e, 
1D-LTE plane-parallel models (ATLAS9+ODFNEW). We also used the package WIDTH9 
for the abundances computation, under some modifications to facilitate the 
handling of the input/output files. Extra IDL\footnote{{\sf IDL} (Interactive 
Data Language) is a registered trademark of ITT Visual Information Solutions.} 
routines were written to optimize the calculation through the spectroscopic 
requirements mentioned before. First, the program computes the atmospheric 
model correspondent to the values of $T_\mathrm{eff}$, $\log~g$, [Fe/H], and 
$\xi$ given as initial guess. The code optimizes one parameter at a time, 
always checking if the optimization of the previous parameter is still valid. 
If the previous optimization is still satisfied the code moves forward to the 
next parameter. If not, the code returns to the previous parameter and 
recalculates the optimization. The final set of spectroscopic parameters is 
obtained once all parameters are optimized consistently. An example of the 
convergence of M2 is shown in Fig.~\ref{fig:m2} for the star HIP~20205, and 
for both ASPL and MASH lists.

The internal errors of the M2 set of spectroscopic stellar parameters were 
obtained as follows: for the $T_\mathrm{eff}$, we changed the slope of the 
linear regression in the [Fe/H] versus $\chi$ diagram by its own 1$\sigma$ 
error. The error in temperature is the difference between this new temperature 
and the previous best value. Similarly, for $\xi$, we applied the same idea, 
changing the slope of the linear regression in the [Fe/H] versus 
$\log(EW$/$\lambda)$ diagram. The $\log~g$ error is estimated by changing this 
parameter until we obtain a difference between the \ion{Fe}{i} or \ion{Fe}{ii} 
abundances that is equal to the larger of their dispersions. For the 
metallicity, we adopted the standard deviation of the \ion{Fe}{i} abundance 
distribution. Table~\ref{tab:m2} shows the stellar parameters derived using M2 
for ASPL and MASH lists.

The metallicities derived by this method agree between giants and dwarfs, 
within the presented uncertainties. However, the internal metallicity uncertainties of 
M2 are relatively larger than for M1, when compared within the same line 
list. On average, the internal uncertainties of M2 are about $\sim$~0.10~dex 
when using the MASH list and $\sim$~0.06 with the ASPL list. The difference in 
the metallicity scale obtained for both lists are also larger when using M2. 
These differences are more significant for the giants and for the hottest 
dwarf of the sample. This result reflects how the line lists were assembled. 
The MASH line list was selected using the Sun as a proxy and has a larger 
number of transitions. This reduces the number of well-isolated lines, free 
from neighboring features, in the spectra of the giants. On the other hand, 
the ASPL list was chosen to have only isolated and uncontaminated transitions 
in the spectra of giants, which consequently reduced the number of lines used 
in the analysis. We discuss the quality of the line lists again in 
Section~\ref{sec:discussion}. On average, the difference between the 
atmospheric parameters obtained with the MASH and ASPL lists for M2 is +68$\pm$92~K 
for $T_\mathrm{eff}$, +0.12$\pm$0.24 for $\log~g$, $-$0.13$\pm$0.21 for $\xi$ 
and +0.09$\pm$0.08 for [Fe/H]. These values are not significantly different from typical 
errors of spectroscopic analyses and from typical comparisons between 
multiple analyses methods \citep[see, e.g.,][]{Hinkel14,Smiljanic14}. 

\begin{sidewaystable*}
\caption[Abundances according to M2.]{Stellar parameters and iron abundances for giants and dwarfs derived according to method M2. The left side of the table shows the results obtained with the MASH line list and the right side the results obtained with the ASPL line list.}\label{tab:m2}
\centering
\ra{1.1}
\small
\begin{tabular}{@{}c c c c c c c c   c c c c c c c c c c c c c c c c c c c c c c@{}}\toprule[1.0pt]
&  \multicolumn{7}{c}{\bf{MASH}}  &  \multicolumn{7}{c}{\bf{ASPL}} \\
\midrule
HIP & $T_\mathrm{eff}$ &  $\log~g$ &  $\xi$ & log$\epsilon_{\mathrm{FeI}}$  & $\mathrm{N(FeI)}$ &  log$\epsilon_{\rm{FeII}}$   & $\mathrm{N(FeII)}$ & $T_\mathrm{eff}$ &  $\log~g$ &  $\xi$ & log$\epsilon_{\mathrm{FeI}}$  & $\mathrm{N(FeI)}$ &  log$\epsilon_{\rm{FeII}}$   & $\mathrm{N(FeII)}$ \\
\cmidrule(r){2-8}\cmidrule(l){9-15}
20205        & 4914$\pm$109  & 2.88$\pm$0.07 & 1.34$\pm$0.16  & 7.62$\pm$0.11 & 25 & 7.62$\pm$0.07 & 10  & 4875$\pm$23 & 2.71$\pm$0.06 & 1.43$\pm$0.06 & 7.53$\pm$0.07 & 21 & 7.52$\pm$0.06 & 7 \\
20455        & 5010$\pm$162 & 3.03$\pm$0.09 & 1.15$\pm$0.21 & 7.72$\pm$0.13 & 24 & 7.72$\pm$0.11 & 14  & 4816$\pm$61 & 2.55$\pm$0.07 & 1.35$\pm$0.05 & 7.50$\pm$0.06 & 19 & 7.51$\pm$0.07 & 7 \\ 
20889        & 4955$\pm$229 & 3.36$\pm$0.11 & 1.08$\pm$0.23 & 7.86$\pm$0.16 & 25 & 7.86$\pm$0.12 & 12  & 4833$\pm$38 &  2.74$\pm$0.08 & 1.41$\pm$0.06 & 7.59$\pm$0.09 & 19 & 7.59$\pm$0.07 & 6 \\ 
\hline
\noalign{\smallskip}
\it{average}  &                 &      &      & \it\textbf{7.73$\pm$0.12} &&  \it\textbf{7.73$\pm$0.12} &        &      &      &      & \it\textbf{7.54$\pm$0.05} & &   \it\textbf{7.54$\pm$0.04}   \\
\hline
\midrule
HIP & $T_\mathrm{eff}$ &  $\log~g$ &  $\xi$ & log$\epsilon_{\mathrm{FeI}}$  & $\mathrm{N(FeI)}$ &  log$\epsilon_{\rm{FeII}}$   & $\mathrm{N(FeII)}$ & $T_\mathrm{eff}$ &  $\log~g$ &  $\xi$ & log$\epsilon_{\mathrm{FeI}}$  & $\mathrm{N(FeI)}$ &  log$\epsilon_{\rm{FeII}}$   & $\mathrm{N(FeII)}$ \\
\cmidrule(r){2-8}\cmidrule(l){9-15}
18946 & 4815$\pm$141  &  4.56$\pm$0.17 & 0.91$\pm$0.10 & 7.50$\pm$0.08   & 20 & 7.49$\pm$0.25 & 10 & 4813$\pm$104  & 4.78$\pm$0.27 & 0.54$\pm$0.20 & 7.57$\pm$0.11   & 18 & 7.57$\pm$0.34 & 7  \\
13976 & 4900$\pm$87  &  4.31$\pm$0.09 & 0.70$\pm$0.19 & 7.60$\pm$0.07   & 20 & 7.59$\pm$0.12 & 9 & 4915$\pm$42  & 4.44$\pm$0.08 & 0.63$\pm$0.17 & 7.58$\pm$0.06   & 20 & 7.58$\pm$0.08 & 5  \\
19098 & 5120$\pm$75  &  4.54$\pm$0.06 & 1.00$\pm$0.19 & 7.53$\pm$0.07   & 21 & 7.53$\pm$0.06 & 9 & 5025$\pm$62  & 4.35$\pm$0.06 & 0.93$\pm$0.13 & 7.51$\pm$0.08   & 25 & 7.53$\pm$0.07 & 7  \\
16529 & 5090$\pm$86  &  4.22$\pm$0.18 & 0.83$\pm$0.13 & 7.55$\pm$0.10   & 23 & 7.55$\pm$0.11 & 11 & 5100$\pm$48  & 4.30$\pm$0.06 & 0.83$\pm$0.10 & 7.53$\pm$0.05   & 20 & 7.53$\pm$0.05 & 6  \\
19934 & 5330$\pm$59  &  4.57$\pm$0.07 & 0.81$\pm$0.14 & 7.62$\pm$0.05  & 21 & 7.60$\pm$0.08 & 10  & 5200$\pm$45  & 4.35$\pm$0.08 & 0.81$\pm$0.11 & 7.52$\pm$0.06   & 25 & 7.52$\pm$0.08 & 7  \\
20130 & 5590$\pm$104  &  4.73$\pm$0.12 & 0.98$\pm$0.17 & 7.62$\pm$0.02   & 17 & 7.61$\pm$0.04 & 8 & 5505$\pm$40  & 4.56$\pm$0.06 & 1.01$\pm$0.07 & 7.55$\pm$0.05   & 25 & 7.54$\pm$0.05 & 6  \\
20146 & 5500$\pm$92  &  4.34$\pm$0.10 & 0.58$\pm$0.12 & 7.68$\pm$0.11   & 27 & 7.68$\pm$0.12 & 11 & 5540$\pm$40  & 4.48$\pm$0.05 & 0.96$\pm$0.08 & 7.57$\pm$0.06   & 27 & 7.58$\pm$0.04 & 6  \\
19781 & 5695$\pm$53  &  4.63$\pm$0.08 & 0.78$\pm$0.12 & 7.66$\pm$0.03   & 17 & 7.66$\pm$0.06 & 8 & 5625$\pm$41  & 4.46$\pm$0.05 & 0.96$\pm$0.07 & 7.56$\pm$0.05   & 27 & 7.57$\pm$0.03 & 6  \\
19793 & 5790$\pm$88  &  4.47$\pm$0.05 & 0.80$\pm$0.20 & 7.66$\pm$0.08   & 28 & 7.65$\pm$0.04 & 7 & 5710$\pm$34  & 4.32$\pm$0.07 & 1.13$\pm$0.07 & 7.54$\pm$0.05   & 26 & 7.53$\pm$0.05 & 7  \\
20899 & 5855$\pm$105  &  4.39$\pm$0.12 & 0.85$\pm$0.17 & 7.62$\pm$0.07   & 28 & 7.62$\pm$0.05 & 8 & 5885$\pm$47  & 4.43$\pm$0.05 & 1.09$\pm$0.07 & 7.58$\pm$0.05   & 26 & 7.58$\pm$0.01 & 5 \\
19148 & 5985$\pm$91  &  4.46$\pm$0.07 & 0.75$\pm$0.15 & 7.62$\pm$0.09   & 30 & 7.63$\pm$0.05 & 9 & 5970$\pm$50  & 4.43$\pm$0.05 & 1.09$\pm$0.07 & 7.55$\pm$0.05   & 26 & 7.55$\pm$0.02 & 6 \\
22422 & 6029$\pm$104  &  4.30$\pm$0.08 & 0.57$\pm$0.28 & 7.73$\pm$0.10   & 32 & 7.73$\pm$0.10 & 11 & 5975$\pm$53  & 4.40$\pm$0.06 & 0.94$\pm$0.09 & 7.62$\pm$0.06   & 30 & 7.62$\pm$0.02 & 6 \\
21112 & 6135$\pm$113  &  4.30$\pm$0.07 & 0.83$\pm$0.24 & 7.66$\pm$0.13   & 36 & 7.66$\pm$0.07 & 11  & 6095$\pm$44  & 4.31$\pm$0.04 & 1.16$\pm$0.09 & 7.55$\pm$0.04   & 24 & 7.55$\pm$0.03 & 7 \\
22566 & 6343$\pm$152  &  4.56$\pm$0.11 & 1.24$\pm$0.14 & 7.69$\pm$0.14   & 33 & 7.69$\pm$0.13 & 11  & 6010$\pm$59  & 4.07$\pm$0.08 & 1.16$\pm$0.09 & 7.55$\pm$0.05   & 24 & 7.56$\pm$0.08 & 6 \\
\hline
\noalign{\smallskip}
\it{average}  &                 &      &      & \it\textbf{7.62$\pm$0.06} & &   \it\textbf{7.62$\pm$0.07} &        &      &      &      & \it\textbf{7.56$\pm$0.03} & &   \it\textbf{7.56$\pm$0.03}   \\
\bottomrule
\end{tabular}
\end{sidewaystable*}

\subsection{Solar abundances and differential analysis}

We also computed the solar metallicity using our solar proxy spectra. This is 
useful to understand the behavior of the metallicity scale derived with 
methods M1 and M2. As mentioned in Section~\ref{sec:data}, for the comparison 
with the giant stars we used the solar reflected spectrum of Ganymede, 
hereafter Sun HARPS. For the comparison with the dwarf stars, we used the 
solar spectrum observed with UVES, hereafter Sun UVES. We decided to have two 
solar proxies to avoid any inconsistencies that may arise from the use of two 
different spectrographs and, therefore, different spectral resolution, 
instrumental profiles, or possible scattered light influence that affects a 
particular spectrograph.

Solar abundances were derived applying the same two methods presented before 
and for both MASH and ASPL lists. For M1, in particular, we fixed the solar 
atmospheric parameters to 5777/4.44/0.90 instead of ranging them within its 
expected errors. For the Sun, we prefer to keep fixed these parameters since 
its errors are too small to produce a noticeable difference in the metallicity 
determination. A 2$\sigma$ clipping of the lines was also applied. For M2, 
the solar parameters were computed exactly as describe in Section~\ref{sec:m2}.

Table~\ref{tab:sun} shows the abundances of \ion{Fe}{i} and \ion{Fe}{ii} for 
our solar proxies, according to M1 and M2, using the ASPL and MASH lists. 
The solar atmospheric parameters found through M2 are very similar to the 
canonical values adopted as fixed in M1. The mean differences between M1 and 
M2 for the solar parameters are 41~K for $T_\mathrm{eff}$, with minimum and 
maximum values of 16~K and 52~K; 0.01~dex for $\log~g$, with minimum and 
maximum values of 0.01~dex and 0.02~dex; and 0.09~$\mathrm{km~s^{-1}}$ for 
$\xi$, with minimum and maximum values of 0.01~$\mathrm{km~s^{-1}}$ and 
0.18~$\mathrm{km~s^{-1}}$. 

Finally, we used the solar values of Table~\ref{tab:sun} as a reference and 
derived the abundances of \ion{Fe}{i} and \ion{Fe}{ii} with respect to the 
Sun for the Hyades stars. The results of this differential analysis are shown 
in Table~\ref{tab:m1m2diff}, for M1 and M2 and for both ASPL and MASH lists. 
Also given are the average metallicities obtained for giants and dwarfs 
according to each methodology. The internal dispersions presented in the table 
were obtained by the squared sum of the internal uncertainty relative to the 
\ion{Fe}{} abundance of the star and the Sun. 

\begin{table*}[!htbp]
 \caption[Solar abundances obtained with M1 and M2 for both MASH and ASPL lists.]{Solar abundances of \ion{Fe}{i} and \ion{Fe}{ii} obtained with M1 and M2 for both MASH and ASPL lists.}\label{tab:sun}
\small
 \label{t:4.7}
\begin{center}
\begin{tabular}{@{}c c c c c c c c c c@{}}\toprule[1.0pt]
M1 &  \multicolumn{3}{c}{\bf{MASH}}  &  \multicolumn{3}{c}{\bf{ASPL}} \\
\midrule
  & log$\epsilon_{\mathrm{FeI}}$  & $\mathrm{N(FeI)}$ &  log$\epsilon_{\rm{FeII}}$   & $\mathrm{N(FeII)}$  & log$\epsilon_{\mathrm{FeI}}$  & $\mathrm{N(FeI)}$ &  log$\epsilon_{\rm{FeII}}$   & $\mathrm{N(FeII)}$ \\
\cmidrule(r){2-5}\cmidrule(l){6-9}
Sun UVES  & 7.46$\pm$0.10 & 36 & 7.45$\pm$0.05 & 14 & 7.45$\pm$0.06 & 31 & 7.42$\pm$0.02 & 5 \\
Sun HARPS  & 7.45$\pm$0.07 & 37 & 7.44$\pm$0.06 & 14 & 7.45$\pm$0.05 & 34 & 7.43$\pm$0.02 & 7\\
\midrule
M2 &  \multicolumn{3}{c}{\bf{MASH}}  &  \multicolumn{3}{c}{\bf{ASPL}} \\
\midrule
  & log$\epsilon_{\mathrm{FeI}}$  & $\mathrm{N(FeI)}$ &  log$\epsilon_{\rm{FeII}}$   & $\mathrm{N(FeII)}$  & log$\epsilon_{\mathrm{FeI}}$  & $\mathrm{N(FeI)}$ &  log$\epsilon_{\rm{FeII}}$   & $\mathrm{N(FeII)}$ \\
\cmidrule(r){2-5}\cmidrule(l){6-9}
Sun UVES  & 7.47$\pm$0.05 & 22 & 7.48$\pm$0.08 & 15 & 7.41$\pm$0.03 & 25 & 7.41$\pm$0.05 & 6 \\
Sun HARPS  & 7.50$\pm$0.05 & 26 & 7.49$\pm$0.07 & 14 & 7.42$\pm$0.05 & 33 & 7.42$\pm$0.02 & 5\\
\bottomrule
\end{tabular}
\end{center}
\end{table*}

\begin{table*}
\caption{Metallicity, with respect to the Sun, derived using M1 and M2.}\label{tab:m1m2diff}
\centering
\ra{1.1}
\small
\begin{tabular}{@{}c c c c c c c c c c@{}}\toprule[1.0pt]
&  \multicolumn{4}{c}{\bf{M1}}  &  \multicolumn{4}{c}{\bf{M2}} \\
\midrule
&  \multicolumn{2}{c}{\bf{MASH}}  &  \multicolumn{2}{c}{\bf{ASPL}} & \multicolumn{2}{c}{\bf{MASH}}  &  \multicolumn{2}{c}{\bf{ASPL}}\\
\midrule
HIP & [FeI/H] &  [FeII/H] &  [FeI/H] & [FeII/H]  & [FeI/H] &  [FeII/H] &  [FeI/H] & [FeII/H]\\
\cmidrule(r){2-3}\cmidrule(l){4-5} \cmidrule(l){6-7}\cmidrule(l){8-9}

20205        & 0.17$\pm$0.12  &  0.18$\pm$0.11   &  0.14$\pm$0.08  &  0.16$\pm$0.06  & 0.12$\pm$0.12  & 0.13$\pm$0.10    &  0.11$\pm$0.09  &  0.10$\pm$0.06 \\
20455        & 0.14$\pm$0.12  &  0.18$\pm$0.11   &  0.11$\pm$0.09  &  0.13$\pm$0.06  & 0.22$\pm$0.14  & 0.23$\pm$0.13    &  0.08$\pm$0.09  &  0.09$\pm$0.07 \\
20889        & 0.25$\pm$0.11  &  0.26$\pm$0.12   &  0.18$\pm$0.09  &  0.20$\pm$0.06  & 0.36$\pm$0.17  & 0.37$\pm$0.14    &  0.17$\pm$0.10  &  0.17$\pm$0.07 \\
\hline
\noalign{\smallskip}
{\bf \emph{average - giants} }       & {\bf 0.19$\pm$0.07}  & {\bf 0.21$\pm$0.05}    & {\bf 0.14$\pm$0.03}  &  {\bf 0.16$\pm$0.03} & {\bf 0.23$\pm$0.12}  & {\bf 0.24$\pm$0.12}    & {\bf 0.12$\pm$0.04}  &  {\bf 0.12$\pm$0.04} \\
\hline
\midrule

18946        & 0.17$\pm$0.13  &  0.18$\pm$0.15   &  0.11$\pm$0.11  &  0.14$\pm$0.21  & 0.03$\pm$0.09  & 0.01$\pm$0.26    &  0.16$\pm$0.11  &  0.16$\pm$0.34 \\
13976        & 0.23$\pm$0.12  &  0.18$\pm$0.12   &  0.19$\pm$0.10  &  0.22$\pm$0.08  & 0.13$\pm$0.09  & 0.11$\pm$0.14    &  0.17$\pm$0.07  &  0.17$\pm$0.09 \\
19098        & 0.16$\pm$0.13  &  0.17$\pm$0.08   &  0.18$\pm$0.11  &  0.21$\pm$0.06  & 0.06$\pm$0.09  & 0.05$\pm$0.10    &  0.10$\pm$0.08  &  0.12$\pm$0.09 \\
16529        & 0.18$\pm$0.14  &  0.19$\pm$0.09   &  0.17$\pm$0.10  &  0.20$\pm$0.05  & 0.08$\pm$0.11  & 0.07$\pm$0.14    &  0.12$\pm$0.06  &  0.12$\pm$0.07 \\
19934        & 0.18$\pm$0.12  &  0.19$\pm$0.08   &  0.17$\pm$0.08  &  0.17$\pm$0.08  & 0.15$\pm$0.07  & 0.12$\pm$0.11    &  0.11$\pm$0.07  &  0.11$\pm$0.09 \\
20130        & 0.20$\pm$0.13  &  0.21$\pm$0.09   &  0.17$\pm$0.08  &  0.17$\pm$0.07  & 0.15$\pm$0.05  & 0.13$\pm$0.09    &  0.14$\pm$0.06  &  0.13$\pm$0.07 \\
20146        & 0.23$\pm$0.12  &  0.24$\pm$0.09   &  0.17$\pm$0.09  &  0.17$\pm$0.04  & 0.21$\pm$0.12  & 0.20$\pm$0.14    &  0.16$\pm$0.07  &  0.17$\pm$0.06 \\
19781        & 0.23$\pm$0.13  &  0.24$\pm$0.09   &  0.15$\pm$0.08  &  0.18$\pm$0.05  & 0.19$\pm$0.06  & 0.18$\pm$0.10    &  0.15$\pm$0.06  &  0.16$\pm$0.06 \\
19793        & 0.27$\pm$0.13  &  0.28$\pm$0.10   &  0.19$\pm$0.10  &  0.22$\pm$0.05  & 0.19$\pm$0.09  & 0.17$\pm$0.09    &  0.13$\pm$0.06  &  0.12$\pm$0.07 \\
20899        & 0.20$\pm$0.12  &  0.21$\pm$0.09   &  0.16$\pm$0.10  &  0.19$\pm$0.04  & 0.15$\pm$0.09  & 0.14$\pm$0.09    &  0.17$\pm$0.06  &  0.17$\pm$0.05 \\
19148        & 0.15$\pm$0.12  &  0.16$\pm$0.10   &  0.17$\pm$0.09  &  0.20$\pm$0.04  & 0.14$\pm$0.10  & 0.15$\pm$0.09    &  0.14$\pm$0.06  &  0.14$\pm$0.05 \\
22422        & 0.20$\pm$0.13  &  0.21$\pm$0.09   &  0.18$\pm$0.08  &  0.21$\pm$0.02  & 0.26$\pm$0.11  & 0.25$\pm$0.13    &  0.21$\pm$0.06  &  0.21$\pm$0.09 \\
21112        & 0.14$\pm$0.13  &  0.18$\pm$0.06   &  0.15$\pm$0.08  &  0.18$\pm$0.04  & 0.19$\pm$0.14  & 0.18$\pm$0.11    &  0.14$\pm$0.07  &  0.14$\pm$0.05 \\
22566        & 0.23$\pm$0.13  &  0.24$\pm$0.10   &  0.22$\pm$0.10  &  0.25$\pm$0.07  & 0.22$\pm$0.15  & 0.21$\pm$0.15    &  0.14$\pm$0.06  &  0.15$\pm$0.09 \\

\hline
\noalign{\smallskip}
{\bf \emph{average - dwarfs} }       & {\bf 0.20$\pm$0.04}  & {\bf 0.21$\pm$0.03}    & {\bf 0.17$\pm$0.02 } & {\bf 0.19$\pm$0.03}  & {\bf 0.15$\pm$0.06}  & {\bf 0.14$\pm$0.06}    & {\bf 0.15$\pm$0.03 } & {\bf 0.15$\pm$0.03} \\
\bottomrule
\end{tabular}
\end{table*}

\section{Results and discussion}\label{sec:discussion}

\subsection{The quality of the EWs}

One step in the spectral analysis that has a substantial influence on the 
final parameters and abundance is the measurement of the EWs. In this work, 
EWs were measured with the automatic code ARES. This code been used on many 
abundance analyses in the literature \citep[e.g.,][]{Adibekyan12,Tabernero12}. 
Such kind of automatic codes are a fast and systematic way to compute EWs that 
minimize the subjectivity of manual measurements using, for example, IRAF. 
Nevertheless, it has been shown in the analysis of the Gaia-ESO Survey spectra 
that different groups measuring EWs in the same spectra, using the same code, 
can still find considerably different values \citep[see the discussion in 
Section 6 of][]{Smiljanic14}. This is the case mainly for two reasons: 
i) there are still some crucial free parameters that need to be adjusted for 
the optimal measurement of EWs by automatic codes, in particular, for continuum 
fitting; and ii) quality control of the measured EWs is still important, as 
the codes  just measure all possible EWs without recognizing potential 
local problems (e.g., unrecognized blends). Regarding the continuum placement, 
in general, there is no consensus on the best way to define the continuum, 
which can be done adopting either a local or a global solution.

It is not our aim in this paper to advocate in favor of choosing a global or 
local continuum normalization. We just stress that regardless of the type of 
normalization that was chosen, care is needed to evaluate whether the 
continuum solution was of good quality or whether it is affecting the measured 
EWs in a negative way. As ARES adopts a local continuum normalization, lines 
in crowded regions of the spectrum can have their EWs underestimated because 
of a low local continuum solution. This effect can become important in the 
spectra of giants, where many lines are stronger than in the spectra of 
dwarfs. Of course, if the line list is composed solely of well-isolated 
spectral features, the local and the global continuum solution should give 
similar results. However, this is not always the case, especially when 
analyzing stars in a wide range of $T_{\rm eff}$. For this reason, as stated 
before, we were careful to visually inspect and exclude measurements that 
could be underestimated because of continuum misplacement. This was 
particularly important in the MASH list, where the lines were basically chosen 
in the solar spectrum and, consequently, this list has spectral features that 
can be blended in the spectra of giants. 

We find that the EWs obtained with IRAF are systematically higher than 
the EWs obtained with ARES. We attribute this behavior mainly to the 
different continuum normalization. The larger discrepancies were found for hot 
dwarfs and for the giants. If we compare the EWs of all transitions, without 
excluding those with suspicious normalization, the differences between IRAF 
and ARES reach up to 13.6~m\AA~for the hot star HIP~22566 and up to 
7.4~m\AA~for the giant HIP~20205. The use of all these measurements would 
certainly lead to an extra source of uncertainties. After the visual 
inspection and exclusion, the difference in EWs causes an effect on the 
abundances that is smaller than 0.05~dex (see Section~\ref{sec:ews}). This is 
similar to typical errors found in abundance analyses that are not line-by-line
 differential.

We also noticed some difficulties of measuring EWs with ARES in hot stars 
where rotational broadening becomes to be important. After several tests with 
different ARES input options, we decided to limit our sample to those stars 
where we were confident that the EWs were well measured with this automatic
 tool. Therefore, we do not include stars hotter than $\sim$6300~K. 

Finally, we also tested the influence of using spectra obtained with two 
different instruments on the EWs scale. To this end, we used a spectrum of the 
giant HIP~20455 observed with UVES, under the same configuration set as the 
dwarfs. We computed metallicities using M1 for the two sets of EWs of this 
star. The mean differences were 0.03~dex for the ASPL list and 0.05~dex for 
the MASH list. Although the difference is slightly higher for the MASH list, 
it is still within the internal uncertainties of this method. Consequently, 
if there is any systematic error in the metallicity introduced by the use of 
different spectrographs, this is comparable to the internal errors of our 
analysis. Therefore, we concluded that the use of spectra from both UVES and 
HARPS instruments should not compromise the metallicity scale between giants 
and dwarfs.

\subsection{Comparison between the line lists}

The use of different line lists has already been reported as one of the 
possible sources of the discrepancies between the metallicity scale of giants 
and dwarfs in open clusters \citep{San2009,San2012}. However, in those works 
only the classical spectroscopic analysis was tested. Here, we are able to 
compared the spectroscopic method (M2) with a set of parameters that is 
independent (method M1). Thus, we can test the performance of the different 
line lists in these different approaches.

When considering the analysis performed with the MASH list, the metallicity 
scale of giants and dwarfs agree better under M1. The difference between the 
average metallicity of giants and dwarfs is $-$0.02~dex with M1 and increases 
to 0.11~dex with M2. This increased difference seems to be driven by the 
giants. The uncertainties of the atmospheric parameters are larger for the 
giants than for the dwarfs (Table~\ref{tab:m2}). Moreover, the metallicity and 
the surface gravity are substantially higher for two of the giants when M2 is 
applied with this list. On the other hand, with the ASPL list there is a 
better agreement between giants and dwarfs in both methods. The difference 
between the average metallicity of giants and dwarfs is $-$0.03~dex for M1 and 
$-$0.02~dex for M2.

We attribute the increased scatter in the metallicity of the giants seen when 
using the MASH list under M2 to the fact the lines were chosen from the solar 
spectrum. For this reason, they are likely more affected by blending features 
that were weak in the Sun, but are significant in the spectrum of a giant 
star. This result indicates that the use of fewer transitions, but that are 
carefully chosen to avoid contaminating blends, is fundamental for an accurate 
determination of atmospheric parameters and of the metallicity scale through 
the spectroscopic method.

We also remark that the improved \emph{gf} values that we adopted for the MASH 
list did not reduce significantly the scatter of the metallicities. Indeed, 
our values of metallicity dispersion are around 0.08-0.10~dex. These are 
similar to the ones reported in \cite{Mashon11}, which are about 
$\sim$~0.09-0.11~dex. Perhaps the reduced number of lines in our analysis is 
not enough to reduce the internal statistical errors even improving the 
\emph{gf} values sources. For the list ASPL, we found an average dispersion on 
the metallicity of around 0.07~dex, which is smaller but still closer to the 
above mentioned values.  

\begin{figure*}[htbp]
\centering
\begin{tabular}{cc}
  \includegraphics[width=0.50\textwidth]{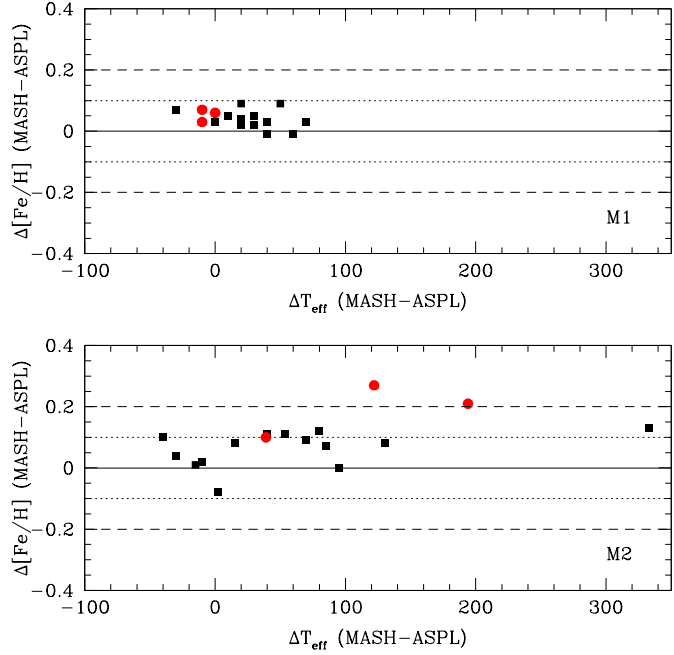} 
\end{tabular}
 \caption{Comparison of the metallicity scale derived with lists MASH  and ASPL as a function of the $T_\mathrm{eff}$ obtained with ASPL list for M1 (upper panel) and M2 (lower panel). In each plot, the null, $\pm$0.1 and $\pm$0.2 abundance differences are represented by the continuous, the dotted and the dashed lines respectively. Red dots refer to the giants while black squares refer to the dwarfs.}\label{fig:linelists}
\end{figure*}

We found a systematic difference of 0.04-0.06~dex between the abundances derived 
using MASH and ASPL lists, and higher for the giants when M2+MASH is applied. The 
metallicities derived with the MASH list are higher than the ones derived with ASPL 
list but also the total average temperature scale is hotter for the MASH list by 
about 20~K when using M1 and by about 70~K when using M2 (see Figure~\ref{fig:linelists}). 
These variations in $T_{\rm eff}$ correspond to an increase in \ion{Fe}{} abundances of 
approx. 0.02-0.06~dex. In addition, the MASH list contains lines which are likely more 
affected by blending, thus we would expect that the cool giants show much larger 
abundances for the MASH list than for the ASPL list, which is the case. Thus, we attribute 
these systematic differences to the higher retrieved $T_{\rm eff}$ and to the presence of 
blended lines in the MASH list. More details about the comparison between M1 and M2 are 
presented in the next section.

As mentioned at the beginning of this subsection, many works have demonstrated how 
abundances (and abundance differences or trends) depend on the adopted line list. 
From our analysis clearly emerges that a proper line list maximises the robustness 
of the results. A similar conclusion has been reached by other authors who recently 
proposed to improve existing line list by customising it to the stars observed, using 
an empirical approach. This method has been applied to cool dwarfs \citep{Tsantaki13} 
and to evolved stars \citep{Adibekyan15}, reaching substantial improvements with 
respect to what obtained with a not optimised line list.

In terms of metallicity dispersion and compatibility of the metallicity scale 
between giants and dwarfs, the ASPL list shows a good performance for both methods 
tested in this work. This is likely related to the selection of well isolated 
line transitions which are equally appropriate for the analysis of giants and 
dwarfs. We, therefore, recommend the use of such list for this kind of 
analysis. 

\subsection{Comparison between the methods}

\begin{figure*}[htbp]
\centering
\begin{tabular}{cc}
\includegraphics[width=0.50\textwidth]{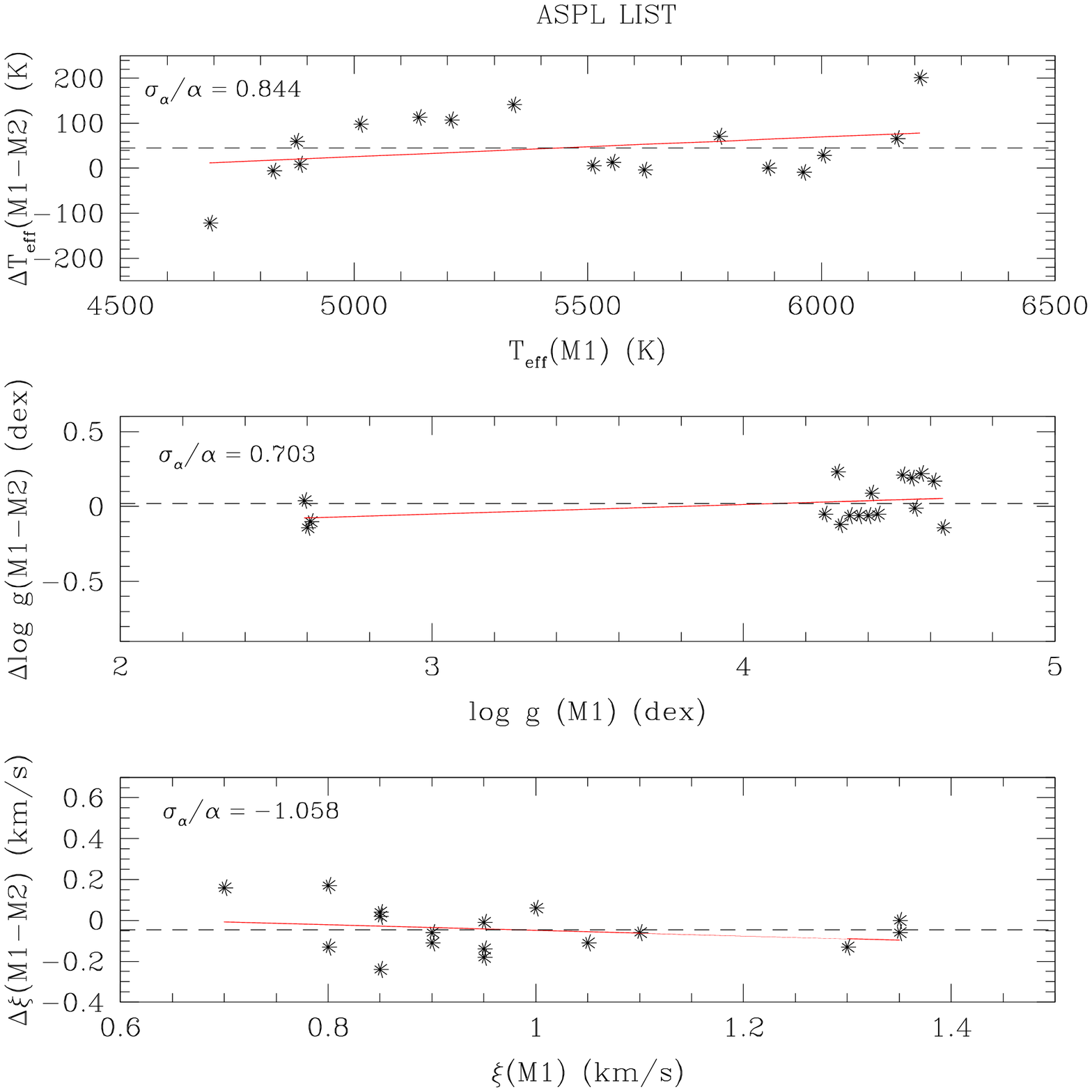} &
\includegraphics[width=0.50\textwidth]{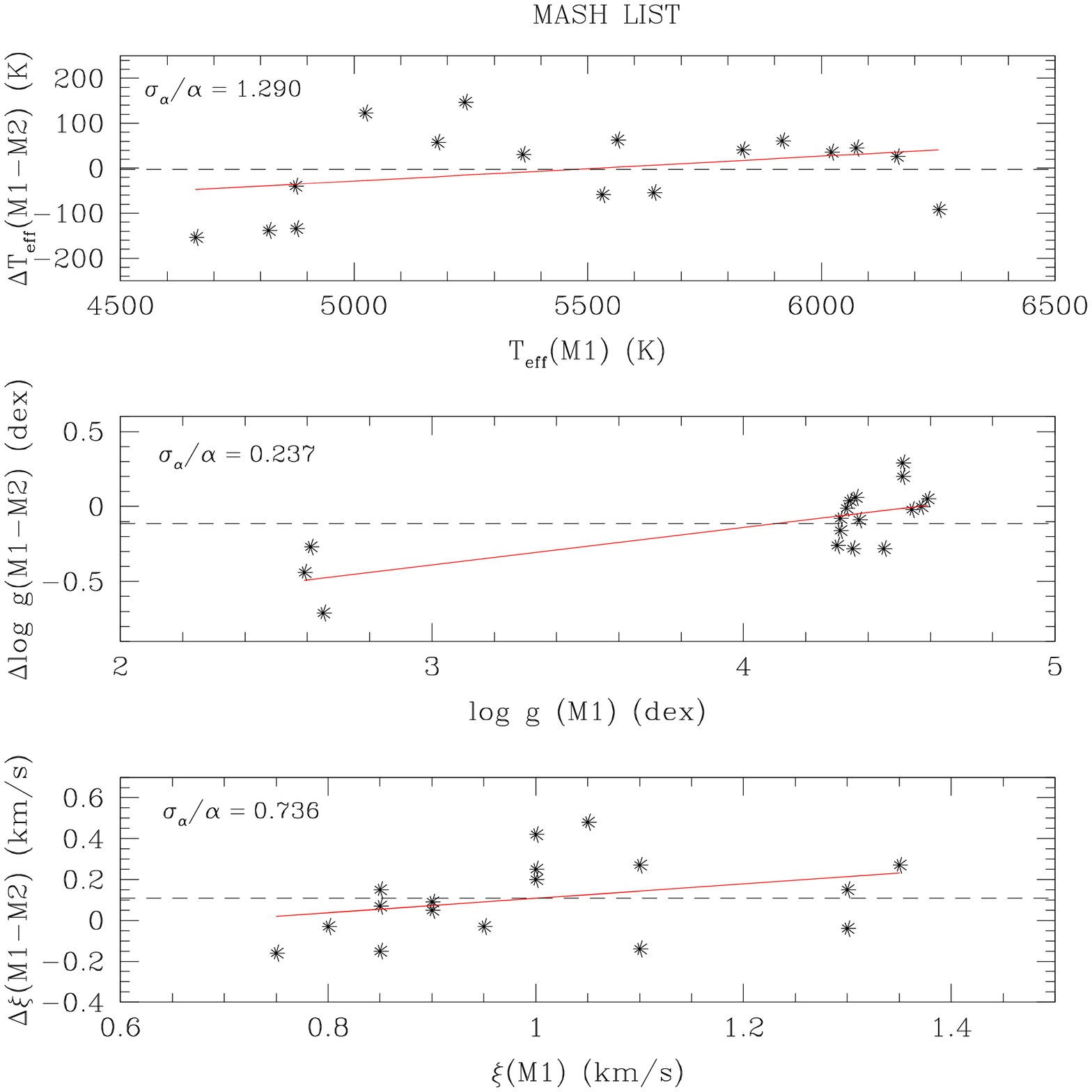}
\end{tabular}
 \caption{Comparison of the atmospheric parameters obtained with methods M1 and M2, as a function of the M1 parameters. The red lines are the linear regressions between the values. \textbf{Left:} parameters obtained with the ASPL line list. \textbf{Right:} parameters obtained with the MASH line list.}\label{fig:methods}
\end{figure*}

In this Section, we compare the performance of methods M1 and M2 to 
investigate possible systematic effects between them. Figure~\ref{fig:methods} 
shows the difference between the atmospheric parameters obtained with M1 and 
M2 as a function of the M1 parameters. For each parameter, we performed a 
linear regression to check for significant trends. These trends would appear 
in case of considerable differences between the atmospheric parameters derived 
with each method. In case of a significant trend, the ratio between the 
slope and its own uncertainty should be less than 0.5, i.e., 
$x$~$>$~2$\sigma_{x}$. 

According to Fig.~\ref{fig:methods}, only $\log~g$ appears significantly 
different between M1 and M2, and only when using the MASH list (see the middle 
plot on the left panel of Fig.~\ref{fig:methods}). On average, the spectroscopy 
surface gravities are larger when using M2 for both line lists, about 
0.10~dex, but in the case of MASH the $\log~g$ values are significantly 
overestimated. This is probably caused by overestimated EWs, due to blending 
contaminants, which result in larger effective temperatures and surface 
gravities. The $\xi$, on the other hand, is higher on average for M1 than for 
M2.

For the ASPL list, the plots shown in Fig.~\ref{fig:methods} do not indicate 
any significant difference between the parameters derived with M1 and M2. 
However, the M1 $T_{\rm eff}$ values are, on average, $\sim$~47~K higher than 
the M2 ones. Although this is a small difference, it is probably responsible 
for the larger metallicities found with M1 in comparison with M2 (in absolute 
values, 0.05~dex for the giants and 0.06~dex for the dwarfs).

The metallicities derived with both line lists show a better agreement when M1 
is applied. This behavior can be more easily noticed by inspecting 
Figure~\ref{fig:linelists}. The differences between the [Fe/H] scale derived 
with the two line lists with M1 are, on average, about $\sim$0.05~dex, and 
constrained within the range of $\pm$0.1~dex (dotted lines) of this Figure. For 
M2, these differences are on average higher about $\sim$0.10~dex, and present a 
larger scatter especially for the giants and for the hottest dwarf HIP~22566. 
Figure~\ref{fig:linelists} shows that a good constraint in the atmospheric 
parameters of a given set of star can produce a consistent metallicity scale for 
giants and dwarfs even when using line lists selected with different approaches, 
not optimized for the analysis of giants.

While the agreement between M1 and M2 is, on average, good, there are a few 
individual stars for which the differences are significant. The 
$T_\mathrm{eff}$ differences are about 100-150~K for the cooler stars 
($T_\mathrm{eff}$~<~5300~K), with both line lists. Such discrepancies for the 
cooler stars do not compromise the overall agreement between the two 
$T_\mathrm{eff}$ scales, but this results are in agreement with 
\cite{Ramirez07} who argue that differences between the spectroscopic and the 
photometric temperature scales are more significant for stars with 
$T_\mathrm{eff}$~$<$~5000~K.  HIP~18946 and HIP~22566 are the stars that 
present the highest discrepancies between the effective temperatures of M1 and 
M2, reaching up to $\sim$~150~K for the MASH list. 

Finally, we tested the compatibility of the giants temperature scale obtained 
with the interferometric angular diameters and the IRFM. This last test was 
done in order too see if, in the absence of angular diameters for the giants, 
the IRFM $T_\mathrm{eff}$ scale would give similar results. As mentioned in 
Sect.~\ref{sec:m1}, the agreement between the IRFM and the interferometric 
$T_\mathrm{eff}$s is excellent. The average absolute difference between them 
is -6$\pm$33~K. This results indicates that, for similar giants, the IRFM 
$T_\mathrm{eff}$ could also be used as input values of temperature in M1.

In general, the metallicities derived using M1 are more consistent between giants 
and dwarfs for the two line lists tested here. This results seems to be a consequence 
of the well constrained set of atmospheric parameters. The metallicities of giants 
and dwarfs are only consistent when derived using M2  with the ASPL list. When using 
M2 associated with the MASH list we found systematic differences especially in the 
giants. Our results confirm that, for the Hyades, the line list is a primary source of 
systematic differences on the metallicity scale of giants and dwarfs. 

\subsection{The Hyades metallicity scale}
 
In this Section we discuss the metallicity scale that we obtained for the 
Hyades. In general, the two methods tested favor  the ASPL over the MASH 
list, as this last one seems to introduce systematic effects on the results, 
in particular for M2. Thus, we consider the results obtained with the ASPL 
list to be of higher quality. There is a good agreement between the 
metallicity of giants and dwarfs when using this list for both methods. 
According to M1 the difference between the average values of the iron 
abundances of giants and dwarfs is 0.03~dex. When M2 is considered, the 
difference between the metallicity of the giants and the dwarfs is 0.02~dex. 
These values are smaller than the internal uncertainties of our methods and 
the typical errors found in classical abundance analysis 
($\sim$~0.05-0.10~dex). Therefore, we consider all the stars together, 
giants and dwarfs, analyzed with ASPL list to revisit the metallicity scale 
of the Hyades.

We combined the average abundances of \ion{Fe}{ii} for giants and dwarfs of 
each list to obtain the metallicity of the cluster. We choose \ion{Fe}{ii} 
abundances because they are safeer abundance indicators than \ion{Fe}{i} 
lines, since \ion{Fe}{i} lines may suffer more of non-LTE effects. We found that 
the average metallicity of the Hyades is 0.18$\pm$0.03~dex or 
0.14$\pm$0.03~dex, according to M1 and M2, respectively,  using the ASPL list. 
These values are compatible with each other within 2$\sigma$. For the MASH 
list, under the M1 analysis, we found a metallicity for the Hyades of 
0.21$\pm$0.04~dex. At this point, we reinforce two main aspects of our work. 
When investigating  giants and dwarfs simultaneously using the classical 
spectroscopic analysis, it is recommended to use a line list especially 
suitable for the spectra of giants. In case the line list was selected based 
on the spectrum of a dwarf star, as the Sun, we recommend that it is more 
appropriate to use a method in the lines of M1. This minimizes the 
systematic effects on the metallicity scale retrieved by the analysis. 

\subsection{Comparison with other works}

We present here a comparison of our results with selected previous analyses of 
Hyades stars: \cite{Paulson03}, \cite{Schuler06a} and \cite{CarPan11}. These 
works were chosen because they have a larger number of stars in common with 
our analysis. As our final recommendation we select the results obtained with 
M1 using the ASPL list. 
 
\cite{Paulson03} analyzed a sample of 90 stars of the Hyades, of which 13 are 
present in our study. Those authors performed a differential spectroscopic 
analysis to derive the atmospheric parameters of the stars. For the sample in 
common, the average difference in $T_{\rm eff}$ between \cite{Paulson03} and 
this work is 58$\pm$59~K. The metallicity reported by \cite{Paulson03} is 
0.13$\pm$0.01~dex, which is within a 2$\sigma$ range of the values found with 
M1 (0.18$\pm$0.03~dex), but it is in excellent agreement with our metallicity 
derived according the same spectroscopic method (M2) of 0.14$\pm$0.03~dex.  

\cite{Schuler06a} analyzed the giants and dwarfs of the Hyades using a 
method similar to M1. The $T_{\rm eff}$ was obtained with the IRFM by 
\cite{Blackwell&Lynas94,Blackwell&Lynas98}. Surface gravities and $\xi$ were 
determined fitting synthetic spectra (1D+LTE) to the observed spectra. For 
the giants, the agreement between this work and our results is remarkably good 
for $\log~g$, $\xi$ and [Fe/H], and reasonable for $T_\mathrm{eff}$. The mean 
differences are 76$\pm$12~K for $T_\mathrm{eff}$, 0.03$\pm$0.06~dex for 
$\log~g$, 0.06$\pm$0.05~$\mathrm{km~s^{-1}}$ for $\xi$ and 0.02$\pm$0.01~dex 
for [Fe/H]. Among the eight dwarfs analyzed by this work, two were also 
included in our analysis. For these objects the agreement between the stellar 
parameters is good for $T_\mathrm{eff}$ and $\log~g$, but the differences 
increase up to 0.40~$\mathrm{km~s^{-1}}$ for $\xi$. The overall metallicity 
for the Hyades recommended by these authors is 0.13~dex, which is also 
compatible within a 2$\sigma$ range of our value retrieved with M1 and ASPL 
list (0.18$\pm$0.03~dex).

Another analysis of the giants of the Hyades was performed by \cite{CarPan11}. 
These authors determined the atmospheric parameters via spectroscopy and 
photometry. The agreement between the set of parameters presented in 
\cite{CarPan11} and our work is reasonably good for $T_\mathrm{eff}$ and good 
for [Fe/H], with average values of about $-$79$\pm$53~K and 
$-$0.03$\pm$0.03~dex, respectively. However, the surface gravities reported by 
these authors are compatible with those that we derived using M2, the 
classical spectroscopic analysis. The differences between the surface 
gravities is likely responsible for the marginal agreement between the 
metallicity scales: 0.11$\pm$0.01~dex derived by \cite{CarPan11} and our 
value of 0.18$\pm$0.03~dex. Although the value of metallicity reported by 
these authors is in excellent agreement with our value of 0.12$\pm$0.04~dex, which 
is the determination of the metallicity for the giants also derived using the 
spectroscopy method (M2 and ASPL list).

\section{Conclusions} \label{sec:conclusion}

We investigated the metallicity scale obtained in a simultaneous analysis 
of dwarfs and giants. Inconsistencies between the metallicities of these 
types of stars have been reported before in the literature. Understanding 
whether these inconsistencies are a real or  systematic effect of the analysis 
methods is important for advancing in a number of different areas of astrophysics. 
Of particular interest is the comparison of metallicities between stars of 
different evolutionary phases hosting planets.

As a test case, we chose a sample of giants and dwarfs in the Hyades open cluster. 
All stars in an open cluster are expected to share the same initial chemical 
composition. They are therefore optimal to test the consistency of analysis methods.

We computed metallicities using two different methods and, for each method, using 
two different line lists. One line list was assembled based on the solar spectrum 
and with the aim of minimizing non-LTE effects in the determination of the 
atmospheric parameters (MASH line list). The other was assembled with the specific 
purpose of analyzing giant stars, and therefore only includes lines that are 
relatively free of blends in the spectra of these objects (ASPL line list).

Analysis method M1 was based on atmospheric parameters, which are independent of 
spectroscopy. Effective temperatures were computed using interferometric angular 
diameters for the giants, and the IRFM for the dwarfs. Surface gravities were 
computed with the help of theoretical evolutionary tracks. For use with M1, we 
presented a new calibration of microturbulence based on 3D hydrodynamical 
models. In this way, one can also use an estimate of $\xi$ that is independent 
of spectroscopic measurements. Method M2 was the classical spectroscopic method 
using \ion{Fe}{i} and \ion{Fe}{ii} lines to constrain the stellar parameters.

We investigate the main steps in the analysis that may affect the metallicity scale 
of dwarfs and giants of the Hyades. Our careful evaluation of EWs has shown that 
differences in the continuum placement can have an effect of about $\sim$ 0.03 dex 
in the precision of the final metallicities. An effect of similar magnitude (0.05~dex) 
can be introduced  using spectra obtained with different instruments. From the 
stand point of the line lists, we found larger systematic differences between the 
metallicity scale for giants and dwarfs using MASH list, where the line selection 
criteria are not optimized for evolved stars. Additionally, we found a systematic 
difference of 0.04-0.06~dex between the two lists used in this work. Lastly, the 
difference on the metallicity scale for the giants and dwarfs using the two different 
methods (M1 and M2) is of about 0.04-0.06~dex, disregarding the results of the giants 
obtained with M2+MASH list. One interesting aspect is that perhaps the use of the 
solar spectrum as a differential reference does not cancel out all these effects. 
Considering all these points, we conclude that the limit of precision of an abundance 
analysis that is not line-by-line strictly differential cannot be better than 
0.03~dex. 

 We show that with a careful determination of atmospheric parameters and a well -selected line list, it is possible to simultaneously analyze giants and dwarfs 
and obtain consistency between their metallicities. When attention to the line list 
is considered, M1 and M2 produce results with a good agreement. The metallicity scale 
of the Hyades obtained with our preferred line list (ASPL) is 0.18$\pm$0.03~dex for M1 
and 0.14$\pm$0.03~dex for M2, and these values are consistent among each other. At the 
very least, it is clear that assembling a line list well suited for the analysis of giants 
is mandatory to obtain consistency between the metallicity scales. When using the 
spectroscopy technique, the metallicity scale of giant and dwarfs using ASPL list agree 
with previous spectroscopic analyses of the Hyades. However, M1 does show a more efficient 
capability to recover the metallicity scale of giants and dwarfs regardless of the line 
list used. The results obtained with M1 for both MASH and ASPL lists are consistent among 
each other, indicating that a set of very well-constrained atmospheric parameters might 
compensate possible systematics from the features of the line list. Therefore, we believe 
that M1 results are more robust.

In the view of our results, we favor as our final metallicity for the stars in Hyades 
open cluster the value of 0.18$\pm$0.03~dex (based on M1 and the ASPL line list). For similar 
studies of giants and dwarfs, in a more general approach, we suggest the use of M1 whenever 
possible. A simultaneous spectroscopic analysis of giants an dwarfs, as done in M2, is only 
recommended under the use of a line list optimized for giant stars.


\begin{acknowledgements}
  L. D. F. acknowledges support through the ESO Studentship Programme, CAPES Programme and CNPq fellowship Programme (166880/2014-0). L. P. acknowledges the Visiting Researcher program of CNPq Brazilian Agency, at the Fed. Univ. of Rio Grande do Norte, Brazil. Support for C.C. is provided by the Ministry for the Economy, Development and Tourism's Programa Iniciativa Cient\'ifica do Mil\^enio through grant IC20009, awarded to the Millennium Institute of Astrophysics (MAS). G.F.P.M. acknowledges financial support by CNPq grant 476909/2006-6, FAPERJ grant APQ1/26/170.687/2004 and ESO, for financial support for a trip to Garching, which enabled participation in this work. We thank Dr. Martin Asplund and Dr. Maria Bergemann for their contribution with the line lists used in this work. This research has made use of the SIMBAD database, operated at CDS, Strasbourg, France, of NASA's Astrophysics Data System, of the WEBDA database, operated at the Department of Theoretical Physics and Astrophysics of the Masaryk University, and of data products from the Two Micron All Sky Survey, which is a joint project of the University of Massachusetts and the Infrared Processing and Analysis Center/California Institute of Technology, funded by the National Aeronautics and Space Administration and the National Science Foundation.
\end{acknowledgements}

\bibliographystyle{aa} 
\bibliography{ldferreira.bib} 

\appendix

\setcounter{table}{0}
\renewcommand{\thetable}{A.\arabic{table}}
\begin{table*}\centering
\ra{1.1}
\small
\caption[LINE LIST DATA MASH.]{Atomic line data for the list MASH and iron abundances according to M1. The data presented here corresponds to all the lines with EW~$\leq$~120mA on the solar spectrum of Ganymede. Some of these lines were excluded as a result of a $\sigma$-clipping to produce the final abundances for the Sun. These are indicated with an $\ast$ symbol in the last column of the table.}\label{tab:mash}
\begin{tabular}{@{}c c c c c c c c@{}}\toprule
Specie & $\lambda$ (\AA) &  $\chi_{\rm{low}}$ (eV) & log~\textit{gf} &  EW (m\AA) &  log$\epsilon_{\rm{Fe}}$ & Clipped\\
\midrule
\multirow{42}{*}{} \ion{Fe}{I} &
  4445.47  &   0.09  &   -5.410  &   40.9        & 7.470  & \\
& 4574.72  &   2.28  &   -2.970  &   54.9        & 7.489  & \\
& 4726.14  &   3.00  &   -3.250  &   18.8        & 7.588  & \\
& 4808.15  &   3.25  &   -2.790  &   26.9        & 7.600  & \\
& 4994.13  &   0.91  &   -3.002  &  104.2        & 7.275  & \\
& 5197.94  &   4.30  &   -1.540  &   36.9        & 7.525  & \\
& 5198.72  &   2.22  &   -2.113  &   97.2        & 7.396  & \\
& 5216.27  &   1.61  &   -2.082  &  117.7        & 7.156  & $\ast$ \\
& 5217.40  &   3.21  &   -1.116  &  111.3        & 7.282  & \\
& 5236.20  &   4.19  &   -1.497  &   32.4        & 7.298  & \\
& 5247.05  &   0.09  &   -4.975  &   66.5        & 7.530  & \\
& 5250.21  &   0.12  &   -4.918  &   64.6        & 7.462  & \\
& 5285.13  &   4.43  &   -1.540  &   27.9        & 7.449  & \\
& 5295.31  &   4.42  &   -1.590  &   29.5        & 7.518  & \\
& 5379.58  &   3.69  &   -1.514  &   61.5        & 7.505  & \\
& 5397.62  &   3.63  &   -2.528  &   28.0        & 7.702  & $\ast$\\
& 5491.83  &   4.19  &   -2.188  &   13.4        & 7.445  & \\
& 5517.06  &   4.21  &   -2.370  &   17.5        & 7.791  & $\ast$\\
& 5522.45  &   4.21  &   -1.450  &   42.9        & 7.463  & \\
& 5607.66  &   4.15  &   -3.437  &   14.6        & 8.699  & $\ast$ \\
& 5638.26  &   4.22  &   -0.770  &   76.1        & 7.394  & $\ast$ \\
& 5662.52  &   4.18  &   -0.573  &   92.8        & 7.402  & \\ 
& 5679.02  &   4.65  &   -0.820  &   59.0        & 7.501  & \\
& 5778.45  &   2.59  &   -3.430  &   22.1        & 7.409  & \\
& 5807.78  &   3.29  &   -3.410  &    8.5        & 7.563  & \\
& 5852.22  &   4.55  &   -1.230  &   39.3        & 7.470  & \\
& 5855.08  &   4.61  &   -1.478  &   23.8        & 7.441  & \\
& 5858.78  &   4.22  &   -2.160  &   12.2        & 7.383  & \\
& 5916.25  &   2.45  &   -2.914  &   54.1        & 7.471  & \\
& 5930.18  &   4.65  &   -0.230  &   86.8        & 7.335  & \\
& 6065.48  &   2.61  &   -1.470  &  116.7        & 7.285  & \\
& 6082.71  &   2.22  &   -3.571  &   33.7        & 7.456  & \\
& 6105.13  &   4.55  &   -2.532  &   12.8        & 8.084  & $\ast$\\
& 6151.62  &   2.18  &   -3.312  &   50.0        & 7.494  & \\
& 6200.31  &   2.61  &   -2.405  &   73.4        & 7.497  & \\
& 6213.43  &   2.22  &   -2.481  &   82.6        & 7.386  & \\
& 6229.23  &   2.85  &   -2.805  &   37.6        & 7.379  & $\ast$\\
& 6252.56  &   2.40  &   -1.727  &  119.8        & 7.437  & \\
& 6421.35  &   2.28  &   -2.020  &  103.7        & 7.326  & \\
& 6481.88  &   2.28  &   -2.985  &   62.4        & 7.497  & \\
& 6518.37  &   2.83  &   -2.373  &   59.0        & 7.361  & \\
& 6608.03  &   2.28  &   -3.930  &   17.6        & 7.428  & \\ \hline
\multirow {15}{*}{} \ion{Fe}{II} &
  4491.40  &   2.84  &   -2.710  &   77.8        & 7.468  & \\
& 4508.29  &   2.84  &   -2.440  &   84.8        & 7.313  & \\
& 4582.83  &   2.83  &   -3.180  &   57.0        & 7.455  & \\
& 4620.52  &   2.82  &   -3.210  &   50.3        & 7.304  & \\
& 5197.58  &   3.22  &   -2.220  &   81.4        & 7.315  & \\
& 5264.81  &   3.22  &   -3.130  &   44.3        & 7.434  & \\ 
& 5284.11  &   2.88  &   -3.195  &   59.9        & 7.525  & \\
& 5414.07  &   3.21  &   -3.580  &   25.6        & 7.414  & \\
& 5425.26  &   3.20  &   -3.220  &   41.5        & 7.429  & \\
& 5991.38  &   3.15  &   -3.647  &   29.5        & 7.506  & \\
& 6239.95  &   3.89  &   -3.573  &   13.8        & 7.635  & $\ast$\\
& 6247.56  &   3.89  &   -2.435  &   52.1        & 7.512  & \\
& 6369.46  &   2.89  &   -4.110  &   18.6        & 7.407  & \\
& 6432.68  &   2.89  &   -3.570  &   41.3        & 7.442  & \\
& 6456.38  &   3.90  &   -2.185  &   61.5        & 7.456  & \\
\bottomrule
\end{tabular}
 \end{table*}

\begin{table*}\centering
\ra{1.1}
\small
\caption[LINE LIST DATA ASPL.]{Atomic line data for the list ASPL and iron abundances according to M1. The data presented here corresponds to all the lines with EW~$\leq$~120mA on the solar spectrum of Ganymede. Lines indicated in the column 3D--$\xi$ list ($\dag$ symbol) were also selected for the microturbulence calibration using 3D models.}\label{tab:aspl}
\begin{tabular}{@{}c c c c c c c@{}}\toprule
Specie & $\lambda$ (\AA) &  $\chi_{\rm{low}}$ (eV) & log~\textit{gf} &  EW (m\AA) &  log$\epsilon_{\rm{Fe}}$ & 3D--$\xi$ list\\
\midrule
\multirow{34}{*}{} \ion{Fe}{I} &
  5044.21  &   2.851  &   -2.058  &     73.9 &  7.396 & \\
& 5225.52  &   0.110  &   -4.789  &     70.9 &  7.482 & $\dag$\\
& 5247.05  &   0.087  &   -4.946  &     66.5 &  7.501 & $\dag$\\
& 5250.21  &   0.121  &   -4.938  &     64.6 &  7.481 & \\
& 5651.47  &   4.473  &   -1.750  &     18.6 &  7.444 & \\
& 5661.35  &   4.284  &   -1.756  &     22.0 &  7.369 & $\dag$\\
& 5679.02  &   4.652  &   -0.750  &     59.0 &  7.431 & $\dag$\\
& 5701.54  &   2.559  &   -2.216  &     83.7 &  7.504 & $\dag$\\
& 5705.46  &   4.301  &   -1.355  &     38.6 &  7.373 & $\dag$\\
& 5793.91  &   4.220  &   -1.619  &     33.3 &  7.435 & $\dag$\\
& 5809.22  &   3.883  &   -1.710  &     50.5 &  7.516 & $\dag$\\
& 5855.08  &   4.608  &   -1.478  &     23.8 &  7.441 & $\dag$\\
& 5956.69  &   0.859  &   -4.605  &     52.3 &  7.532 & $\dag$\\
& 6027.05  &   4.076  &   -1.090  &     63.1 &  7.397 & $\dag$\\
& 6065.48  &   2.609  &   -1.530  &    116.7 &  7.345 & $\dag$\\
& 6093.64  &   4.607  &   -1.300  &     30.8 &  7.395 & $\dag$\\ 
& 6096.67  &   3.984  &   -1.810  &     37.5 &  7.455 & $\dag$\\
& 6151.62  &   2.176  &   -3.299  &     50.0 &  7.481 & $\dag$\\
& 6165.36  &   4.143  &   -1.460  &     44.9 &  7.454 & $\dag$\\
& 6173.33  &   2.223  &   -2.880  &     67.7 &  7.484 & $\dag$\\
& 6200.31  &   2.609  &   -2.437  &     73.5 &  7.532 & $\dag$\\
& 6213.43  &   2.223  &   -2.520  &     82.6 &  7.425 & \\
& 6240.65  &   2.223  &   -3.233  &     48.5 &  7.427 & $\dag$ \\
& 6252.56  &   2.404  &   -1.687  &    119.8 &  7.397 & $\dag$\\
& 6265.13  &   2.176  &   -2.550  &     84.9 &  7.448 & $\dag$\\
& 6270.23  &   2.858  &   -2.540  &     52.2 &  7.430 & $\dag$\\
& 6430.85  &   2.176  &   -2.006  &    110.0 &  7.318 & $\dag$\\
& 6498.94  &   0.958  &   -4.699  &     45.8 &  7.543 & $\dag$\\
& 6703.57  &   2.759  &   -3.023  &     37.7 &  7.488 & $\dag$\\
& 6705.10  &   4.607  &   -0.980  &     46.5 &  7.446 & $\dag$\\
& 6713.75  &   4.795  &   -1.400  &     21.1 &  7.438 & $\dag$\\
& 6726.67  &   4.607  &   -1.030  &     46.7 &  7.499 & $\dag$\\
& 6750.15  &   2.424  &   -2.621  &     74.3 &  7.504 & $\dag$\\
& 6810.26  &   4.607  &   -0.986  &     48.9 &  7.413 & $\dag$\\ \hline
\multirow {7}{*}{} \ion{Fe}{II} &
  5197.58  &   3.231  &   -2.220  &     81.4 &  7.415 & $\dag$\\
& 5234.62  &   3.221  &   -2.180  &     82.8 &  7.388 & $\dag$\\
& 5264.81  &   3.230  &   -3.130  &     44.3 &  7.461 & $\dag$\\
& 5414.07  &   3.221  &   -3.580  &     25.6 &  7.414 & $\dag$\\
& 5425.26  &   3.200  &   -3.220  &     41.5 &  7.453 & $\dag$\\
& 6369.46  &   2.891  &   -4.110  &     18.6 &  7.407 & $\dag$\\
& 6432.68  &   2.891  &   -3.570  &     41.3 &  7.442 & $\dag$\\
\bottomrule
\end{tabular}
 \end{table*}

\end{document}